\documentclass{article}
\usepackage{jheppub}

\pdfoutput=1
\usepackage{amsmath,amssymb,graphicx}
\usepackage{longtable}
\usepackage{verbatim}
\usepackage{float}

\usepackage[usenames,dvipsnames]{xcolor}
\definecolor{darkerblue}{rgb}{0.2,0.2,0.5}

\usepackage{tikz}
\usetikzlibrary{trees}
\usetikzlibrary{decorations.pathmorphing}
\usetikzlibrary{decorations.markings}
\tikzset{
    photon/.style={decorate, decoration={snake}, draw=black},
    wino/.style={draw=redwine},    
    electron/.style={draw=black, postaction={decorate},
        decoration={markings,mark=at position .55 with {\arrow[draw=black]{>}}}},
    scalar/.style={draw=black, dashed,postaction={decorate},
        decoration={markings,mark=at position .55 with {\arrow[draw=black]{>}}}},
    gluon/.style={decorate, draw=black,
        decoration={coil,amplitude=4pt, segment length=5pt}}
}

\usepackage{afterpage}

\newcommand{\dahad}{\Delta \alpha^{(5)}_{\rm had}}

\newcommand{\bear}{\begin{array}}
\newcommand{\ear}{\end{array}}

\newcommand{\beq}{\begin{eqnarray}}
\newcommand{\eeq}{\end{eqnarray}}
\newcommand{\beqa}{\begin{eqnarray}}
\newcommand{\eeqa}{\end{eqnarray}}

\def\OMIT#1{{}}
\newcommand{\lsim}{\mathrel{\rlap{\lower4pt\hbox{\hskip1pt$\sim$}}
    \raise1pt\hbox{$<$}}}         
\newcommand{\gsim}{\mathrel{\rlap{\lower4pt\hbox{\hskip1pt$\sim$}}
    \raise1pt\hbox{$>$}}}         

\begin{document}

\title{\bf \color{darkerblue} Possible Futures of Electroweak Precision:\\ ILC, FCC-ee, and CEPC}

\author[a]{JiJi Fan,} 
\author[b]{Matthew Reece}
\author[c]{and Lian-Tao Wang}
\affiliation[a]{Department of Physics, Syracuse University, Syracuse, NY, 13210, USA}
\affiliation[b]{Department of Physics, Harvard University, Cambridge, MA 02138, USA}
\affiliation[c]{Enrico Fermi Institute and Kavli Institute for Cosmological Physics, \\University of Chicago, Chicago, IL 60637, USA}

\abstract{ 
The future of high-precision electroweak physics lies in $e^+ e^-$ collider measurements of properties of the $Z$ boson, the $W$ boson, the Higgs boson, and the top quark. We estimate the expected performance of three possible future colliders: the ILC, FCC-ee (formerly known as TLEP), and CEPC. In particular, we present the first estimates of the possible reach of CEPC, China's proposed Circular Electron-Positron Collider, for the oblique parameters $S$ and $T$ and for seven-parameter fits of Higgs couplings. These results allow the physics potential for CEPC to be compared with that of the ILC and FCC-ee. We also show how the constraints on $S$ and $T$ would evolve as the uncertainties on each of the most important input measurements change separately. This clarifies the basic physics goals for future colliders. To improve on the current precision,  the highest priorities are improving the uncertainties on $m_W$ and $\sin^2 \theta_{\rm eff}$. At the same time, improved measurements of the top mass, the $Z$ mass, the running of $\alpha$, and the $Z$ width will offer further improvement which will determine the ultimate reach. Each of the possible future colliders we consider has strong prospects for probing TeV-scale electroweak physics.
}
\maketitle

\section{Introduction}
\label{sec:intro}

The discovery of the Higgs boson has ushered in a new era of electroweak physics. The Standard Model has proved to be essentially correct, at least as a low-energy effective field theory, in its description of electroweak symmetry breaking as due to a light, weakly coupled scalar boson. However, the physics giving rise to the Higgs potential remains completely unclear. If there is a small amount of fine-tuning in the Higgs sector, we expect new physics at nearby scales. Perhaps the Higgs is composite (e.g. a pseudo-Nambu Goldstone boson), or perhaps supersymmetry cuts off the quadratic divergence in the Higgs mass. Although the Large Hadron Collider may yet discover new particles that offer clues to these possibilities, precision measurements of electroweak physics including the Higgs boson's properties may also offer powerful probes of electroweak symmetry breaking. Several compelling possibilities for the next step forward in high-precision electroweak physics exist: the International Linear Collider~\cite{Baer:2013cma}, which may be built in Japan; FCC-ee, a future circular collider formerly known as TLEP~\cite{Gomez-Ceballos:2013zzn}; and the CEPC, a new proposal for an electron--positron collider in China (see \href{http://cepc.ihep.ac.cn}{http://cepc.ihep.ac.cn}).

Our goal in this paper is to assess the physics potential of these different colliders, including a first look at CEPC's potential accuracy in measurements of Higgs boson couplings and in fits of the oblique parameters $S$ and $T$~\cite{Peskin:1991sw, Peskin:1990zt} (see also~\cite{Kennedy:1988sn, Holdom:1990tc, Golden:1990ig}). These correspond, in an effective operator language (reviewed in ref.~\cite{Han:2004az,Han:2008es}), to adding to the Lagrangian the following dimension-six operators from the minimal basis of operators~\cite{Grzadkowski:2010es}:
\beq
{\cal L}_{\rm oblique} = S \left(\frac{\alpha}{4 \sin \theta_W \cos \theta_W v^2}\right) h^\dagger W^{i \mu \nu} \sigma^i h B_{\mu \nu} - T \left(\frac{2 \alpha}{v^2}\right) \left|h^\dagger D_\mu h\right|^2, \label{eq:Loblique}
\eeq
where $h$ is the Standard Model Higgs doublet, and we follow the convention $\left<h\right> \approx \frac{v}{\sqrt{2}}$ so that $v \approx 246$ GeV. Integrating out any SU(2)$_L$ multiplet containing states that are split by electroweak symmetry breaking---for instance, the left-handed doublet of stops and sbottoms in a supersymmetric theory---will produce a contribution to $S$. The masses must additionally be split by custodial symmetry-violating effects to contribute to $T$. For example, in the case of the stop and sbottom sector we have both, and $T$ is numerically dominant~\cite{Drees:1991zk}. 

In this paper we estimate the size of the region in the $(S, T)$ plane that will be allowed after several suites of high-precision measurements: a ``GigaZ'' program at the ILC, a ``TeraZ'' program at FCC-ee, extended runs of FCC-ee combining $Z$ pole data with data at the $W^+ W^-$ threshold and the $t{\overline t}$ threshold, and the $Z$ pole program of CEPC. We present a self-contained discussion of many of the relative advantages and disadvantages of the different machines; for example, the $Z$ mass measurement will be improved only at circular colliders, which can follow LEP in exploiting resonant spin depolarization. We also emphasize the basic physics of the fits and their potential bottlenecks, specifying the goals of the electroweak program in future colliders in order to achieve the best sensitivity. For example, given current data the highest priorities are reducing the uncertainties on $m_W$ for determination of $T$ and of $\sin^2 \theta_{\rm eff}$ for determination of $S$, while improved measurements of the top quark mass or the hadronic contribution to the running of $\alpha$ become important only once other error bars have been significantly reduced. We hope that a clear discussion of the physics underlying electroweak fits will help in the planning of future machines, especially for CEPC which is still at a very early stage. In a companion paper, we will apply the results of this paper to assessing the reach of future $e^+ e^-$ colliders for natural SUSY scenarios~\cite{Fan:2014axa}.

Current work on future $e^+ e^-$ colliders draws on an extensive older literature; see, for instance, refs.~\cite{Baur:1996zi,Gunion:1996cn, Heinemeyer:1999zd, Hawkings:1999ac,Erler:2000jg}. For the most part, in determining the expected accuracy achieved by future colliders we will refer to recent review articles, working group reports, and studies for the ILC and TLEP, to which we refer the reader for a more extensive bibliography of the years of studies that have led to the current estimates~\cite{Baer:2013cma,Gomez-Ceballos:2013zzn,Asner:2013psa,Dawson:2013bba,Baak:2013fwa}. Results in our plots labeled ``ILC'' or ``TLEP'' should always be understood to mean the new physics reach assuming the tabulated measurement precisions we have extracted from ILC and TLEP literature (displayed in Tables~\ref{tab:observables} and~\ref{tab:observables2} below). In particular, we are reserving judgment about the relative measurement precision of the machines or about how conservative or optimistic various numbers in the published tables might be. Our results have some overlap with recent work presented by Satoshi Mishima~\cite{MishimaTalk} and Henning, Lu, and Murayama~\cite{Henning:2014gca}.

The paper is organized as follows. In Sec.~\ref{sec:oblique}, we describe the general procedure of the electroweak fit and show the sensitivities of current and future experiments such as ILC and TLEP to new physics that could be encoded in the $S$ and $T$ parameters. In Sec.~\ref{sec:cepc}, we present the first estimate of the reach for new physics of the electroweak program at CEPC and discuss possible improvements for that program. In Sec.~\ref{sec:details}, we explain the details of the uncertainties used in our fits. In Sec.~\ref{sec:todo}, we explain how improving each observable helps with the fit and offer guidelines for the most important steps to take in future electroweak programs. In Sec.~\ref{sec:CEPChiggs}, we estimate the reach of the Higgs measurements at CEPC using a seven-parameter fit. In Sec.~\ref{sec:newphysics}, we discuss the complementarity between the electroweak probes and Higgs probes in new physics reach in two simple examples: composite Higgs theories with Higgs as a pseudo-Nambu-Goldstone boson and SUSY with a light left-handed stop.  We conclude in Sec.~\ref{sec:conclusion}. 

\section{Global Fit of Electroweak Observables with Oblique Corrections}
\label{sec:oblique}
To study the prospects of electroweak precision tests for future LHC upgrades, the ILC/GigaZ, and FCC-ee (formerly known as TLEP)\footnote{We observe that ``Future Circular Collider'' will presumably cease to be the name when the collider is actually built, in which case a new name will have to be found. Perhaps ``TLEP.''}, we find it sufficient to perform an electroweak fit with a simplified set of input observables following the strategy of the Gfitter group~\cite{Baak:2014ora}. The simplified set of observables includes five observables that are free to vary in the fit: the top mass $m_t$, the $Z$ boson mass $m_Z$, the Higgs mass $m_h$, the strong coupling constant at $Z$ pole $\alpha_s(M_Z^2)$ and the hadronic contribution to the running of $\alpha$: $\Delta\alpha_{\rm had}^{(5)}(M_Z^2)$. The remaining three observables, the $W$ boson mass $m_W$, the effective weak mixing angle $\sin^2 \theta_{\rm eff}^{\ell}$ and the $Z$ boson decay width $\Gamma_Z$, are determined by the values of the five free observables in the SM. The SM parametrizations of these three observables based on full two-loop calculations (except for one missing piece in $\Gamma_Z$) could be found in~\cite{Awramik:2003rn, Awramik:2006uz, Freitas:2014hra}. 

Compared to the full fit, the main difference is that in the simplified fit, the $Z$ pole asymmetry observables are summarized into a single value of the effective weak mixing angle $\sin^2 \theta_{\rm eff}^\ell$. This parameter is not measured directly, but inferred from several other observables. The combination of LEP and SLD results relied on six measurements to determine $\sin^2 \theta_{\rm eff}^\ell$: the leptonic forward-backward asymmetry $A^{0,\ell}_{\rm FB}$, ${\cal A}_\ell$ inferred from tau polarization, ${\cal A}_\ell$ from SLD, $A^{0,b}_{\rm FB}$, $A^{0,c}_{\rm FB}$, and the hadronic charge asymmetry $Q^{\rm had}_{\rm FB}$ (see Fig.~7.6 of ref.~\cite{ALEPH:2005ab}). The smallest uncertainties in the individual determinations of $\sin^2 \theta_{\rm eff}^\ell$ were from ${\cal A}_\ell({\rm SLD})$ and $A^{0,b}_{\rm FB}$. The asymmetry parameter for a given fermion is defined as
\beq
{\cal A}_f = \frac{g_{Lf}^2 - g_{Rf}^2}{g_{Lf}^2 + g_{Rf}^2}
\eeq
and can be inferred from forward-backward or left-right asymmetry measurements. Although a future $e^+ e^-$ collider will perform a similar fit to several observables, for our purposes we can focus on the measurement of $A_{\rm LR}$ as a proxy for $\sin^2 \theta_{\rm eff}^\ell$.  This is possible as the asymmetries are related to the effective weak mixing angle in a simple way as 
\beq
A^0_{\rm LR} = {\cal A}_e = \frac{2(1-4\sin^2 \theta_{\rm eff}^{\ell})}{1+(1-4\sin^2 \theta_{\rm eff}^{\ell})^2}.
\eeq
Notice that by this relation, the relative precision of $\sin^2\theta_{\rm eff}^{\ell}$ will be smaller than that of $A_{\rm LR}$ by about an order of magnitude. For instance, the relative error of $A_\ell$ at SLD is $\sim 10^{-2}$, which could be translated to a relative error of $\sin^2\theta_{\rm eff}^{\ell}$ of order $10^{-3}$ (see Sec.~3.1.6 of~\cite{ALEPH:2005ab}). Fans of the Barbieri-Giudice log-derivative tuning measure~\cite{Barbieri:1987fn} may pause to contemplate whether they believe the proximity of the weak mixing angle squared to $1/4$ in the Standard Model corresponds to a factor of 10 fine tuning; we prefer the Potter Stewart measure~\cite{Stewart:1964} and don't see tuning here. 
We will be interested in new physics affecting the oblique parameters $S$ and $T$~\cite{Peskin:1991sw, Peskin:1990zt}. More specifically, the new physics contribution to the electroweak observables can be expressed as a linear function of $S, T$ and $U$~\cite{Peskin:1991sw, Peskin:1990zt, Maksymyk:1993zm, Burgess:1993mg,Burgess:1993vc} (a collection of these formulas could be found in Appendix A of~\cite{Ciuchini:2013pca}). The $U$ parameter is negligible in many new physics scenarios, so we will set it to be zero throughout the analysis. The deviation of all electroweak observables from the SM prediction depends on only three linear combinations of $S$ and $T$: 
\beq
\Delta m_W, \Delta\Gamma_W &\propto & S - 1.54 T \nonumber \\
\Delta\sin^2 \theta_{\rm eff}^{\ell}, \Delta R_\ell, \Delta \sigma_{\rm had}^0& \propto & S - 0.71 T \nonumber \\
\Delta\Gamma_Z &\propto & S-2.76 T.
\eeq
This justifies our choice to use only $m_W, \sin^2 \theta_{\rm eff}^{\ell}$, and $\Gamma_Z$ in the analysis to bound $S$ and $T$ as they suffice to define the ellipse of allowed $S$ and $T$. Notice that the simplified fit of the Gfitter group~\cite{Baak:2014ora} also included $R_\ell$ in addition to $m_W, \sin^2 \theta_{\rm eff}^{\ell}$ and $\Gamma_Z$. We checked that the inclusion doesn't significantly change the result of the fit. 
The set of oblique parameters could be larger beyond the minimal $S$ and $T$~\cite{Barbieri:2004qk}. For instance, there are $Y$ and $W$ related to coefficients of dimension-six current--current operators for hypercharge and SU(2)$_L$. The coefficients of these operators are usually small in typical perturbative theories, so they are less useful than $S$ and $T$ in many cases \cite{Cho:1994yu,Fan:2014axa}. The $U$ parameter is dimension eight and thus is also usually very small. Nonetheless, it could be worthwhile for future studies to include them.

To assess the compatibility of a point in the $(S, T)$ plane with current and future electroweak data, we compute a modified $\chi^2$ function, which takes into account the theory uncertainties with a flat prior,
\beq
\chi^2_{\rm mod} = \sum_j \left[-2 \log \left({\rm erf}\left(\frac{M_j-O_j+\delta_j}{\sqrt{2}\sigma_j}\right) - {\rm erf}\left(\frac{M_j-O_j-\delta_j}{\sqrt{2}\sigma_j}\right)\right)-2\log \left(\sqrt{2\pi} \sigma_j\right)\right],
\eeq
where the index $j$ runs over all the observables in Table~\ref{tab:observables} and Table~\ref{tab:observables2}. $M_j$ is the measured value of the observable $j$. For the convenience of comparison, we will set all $M$'s in every experiment to be the SM central values, which means that the free observables take their current measured values while the derived ones take the current values of the SM predictions. $O_j$ is the predicted value of the observable $j$ in the theory assuming perfect measurement. It is a function of the free parameters in the fit including $S$ and $T$. $\sigma_j$ and $\delta_j$ are the experiment and theory 1$\sigma$ uncertainties respectively. The derivation of this modified $\chi^2$ function could be found in Appendix~\ref{app:theory}. This definition will approach the usual $\chi^2$ function when theory uncertainty goes to zero. It should also be noted that we neglect correlations between the experimental uncertainties in the simplified fit, which we expect to be small.

\subsection{Prospects for Electroweak Precision at the ILC and FCC-ee}
The prospects for electroweak precision at the ILC and FCC-ee have already been presented in~\cite{Baak:2014ora} and a talk by Satoshi Mishima~\cite{MishimaTalk}. In this subsection, we will carry out the simplified fit described above and present our results, which approximately agree with the results in the literature~\cite{Baak:2014ora, MishimaTalk}.
The observables used in the fit with their current values and estimated future precisions for ILC and FCC-ee could be found in Tables~\ref{tab:observables} and~\ref{tab:observables2}. In Sec.~\ref{sec:details}, we will explain in details the origins of all the numbers we used. 

\begin{figure}[!h]
\begin{center}
\includegraphics[width=0.4\textwidth]{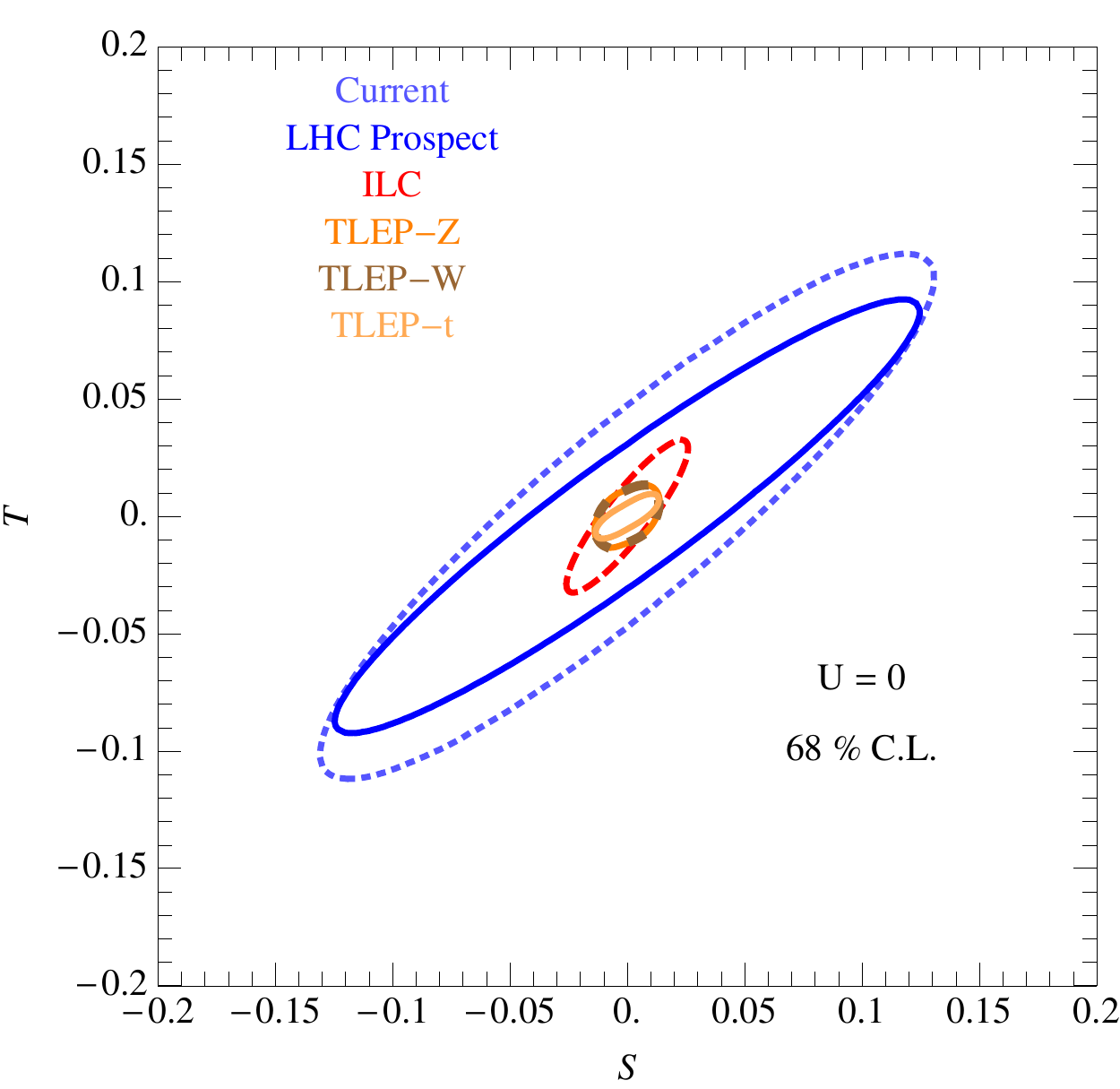} \quad \includegraphics[width=0.4\textwidth]{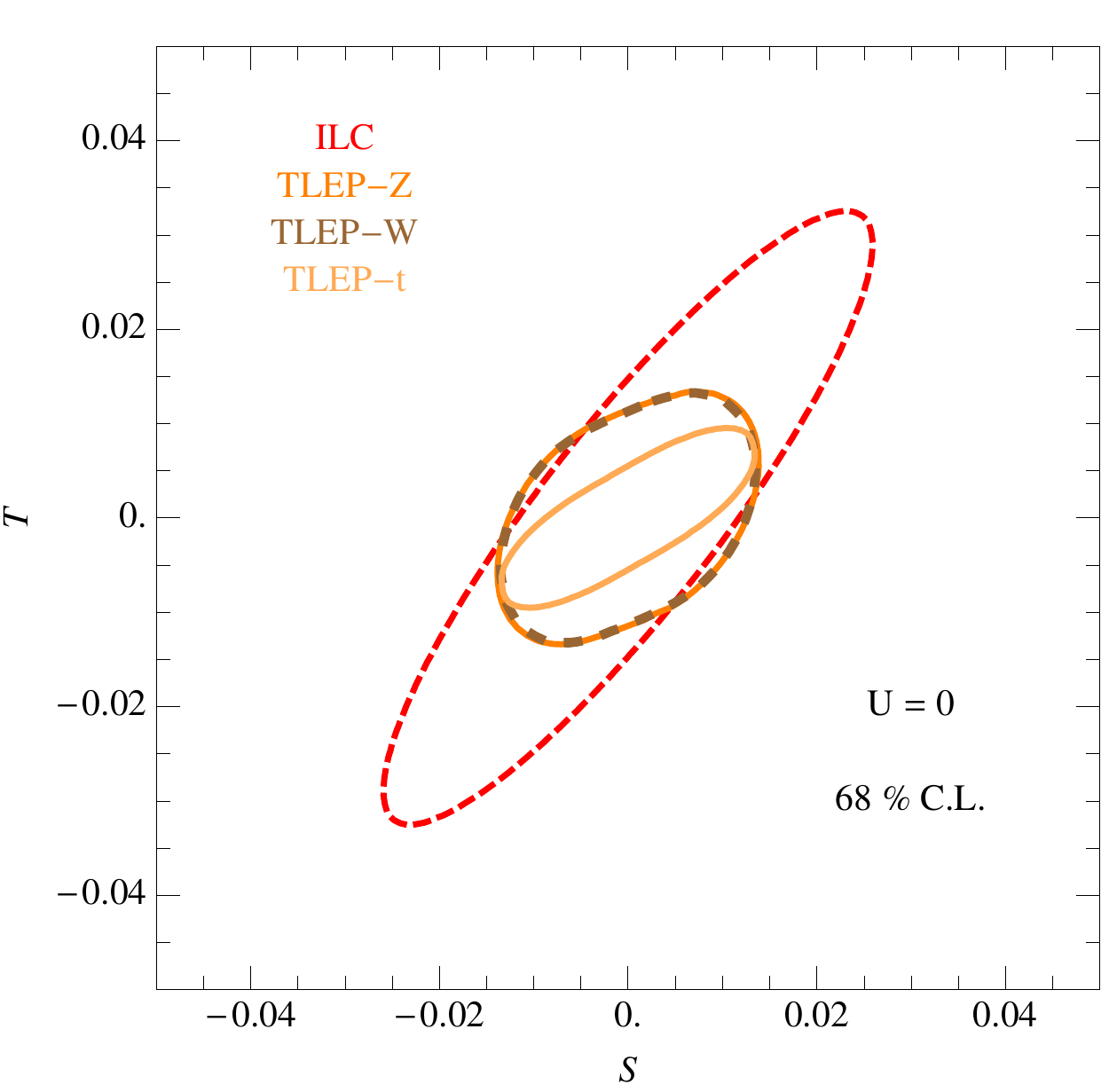}
\end{center}
\caption{Left: 68\% C.L. contours of $S$ and $T$ for different experiments using the simplified fit as described in Tables~\ref{tab:observables} and~\ref{tab:observables2}. Right: a magnified view of 68\% C.L. contours of $S$ and $T$ for ILC and TLEP. We set the best fit point to be $S=T=0$, which corresponds to the current SM values. Our results are in approximate agreement with the current fit from ref.~\cite{pdgreview, Ciuchini:2013pca}, current/LHC14/ILC results by the Gfitter group~\cite{Baak:2014ora}, the TLEP result from a talk by Satoshi Mishima~\cite{MishimaTalk}. The contours of TLEP-$Z$ and TLEP-$W$ almost overlap on top of each other. }
\label{fig:ST}
\end{figure}

We performed profile likelihood fits to map out the allowed $(S,T)$ regions by varying the free electroweak observables in the fit to minimize $\chi^2_{\rm mod}$ for given $S$ and $T$. The boundaries of allowed $S$ and $T$ parameters for different experiments at 68\% C.L. are presented in Fig.~\ref{fig:ST}. Strictly speaking, the best fit point of current data is slightly away from the SM but to facilitate comparisons, we set the best fit points for both current and future data to be at the origin with $S=T=0$, which corresponds to the SM. Currently, the 1$\sigma$ allowed range of $S$ and $T$ is about 0.1 which will be reduced to $\lesssim 0.03$ at ILC, $\lesssim 0.01$ at TLEP. 

\begin{table}[!h]\footnotesize
\centering
\setlength{\tabcolsep}{.3em}
\begin{tabular}{|c|c|c|c|}
\hline
& Present data & LHC14 &  ILC/GigaZ  \\
\hline
$\alpha_s(M_Z^2)$ &
  $0.1185\pm 0.0006$~\cite{Beringer:1900zz} & 
  $\pm 0.0006$ & 
  $\pm 1.0 \times 10^{-4}$~\cite{Lepage:2014fla} 
\\
$\Delta\alpha_{\rm had}^{(5)}(M_Z^2) $ &
  $\left(276.5 \pm 0.8\right) \times 10^{-4} $~\cite{Bodenstein:2012pw} &  
  $\pm 4.7 \times 10^{-5}$~\cite{Baak:2014ora} & 
  $\pm 4.7 \times 10^{-5}$~\cite{Baak:2014ora} 
\\
$m_Z$ [GeV] &
  $91.1875\pm0.0021$~\cite{ALEPH:2005ab} & 
   $\pm0.0021$~\cite{Baak:2014ora}& 
   $\pm0.0021$~\cite{Baak:2014ora} 
\\
$m_t$ [GeV] (pole)&
  $173.34\pm0.76_{\rm exp}$~\cite{ATLAS:2014wva} $\pm 0.5_{\rm th}$~\cite{Baak:2014ora}& 
   $\pm 0.6_{\rm exp} \pm 0.25_{\rm th}$~\cite{Baak:2014ora}& 
    $\pm 0.03_{\rm exp} \pm 0.1_{\rm th}$~\cite{Baak:2014ora} 
\\
$m_h$ [GeV] &
  $125.14 \pm 0.24$~\cite{Baak:2014ora} & 
   $<\pm 0.1$~\cite{Baak:2014ora}& 
    $<\pm 0.1$~\cite{Baak:2014ora} 
\\
\hline
$m_W$ [GeV] &
$80.385\pm 0.015_{\rm exp}$~\cite{Beringer:1900zz}$\pm 0.004_{\rm th}$~\cite{Awramik:2003rn}& 
   $\left(\pm 8_{\rm exp} \pm 4_{\rm th}\right) \times 10^{-3}$~\cite{Baak:2014ora,Awramik:2003rn}& 
    $\left(\pm 5_{\rm exp} \pm 1_{\rm th}\right) \times 10^{-3}$~\cite{Baak:2014ora,Freitas:2013xga} 
   \\
   $\sin^2\theta^{\ell}_{\rm eff}$  &
$(23153 \pm 16)\times 10^{-5}$~\cite{ALEPH:2005ab} & 
      $\pm 16 \times 10^{-5}$    & 
     $\left(\pm 1.3_{\rm exp} \pm 1.5_{\rm th}\right) \times 10^{-5} $~\cite{Baak:2013fwa,Freitas:2013xga}      
   \\
   $\Gamma_{Z}$ [GeV] &
$2.4952\pm 0.0023$~\cite{ALEPH:2005ab} & 
    $\pm 0.0023$      & 
     $\pm 0.001$~\cite{AguilarSaavedra:2001rg}      
   \\
\hline
\end{tabular}
\caption{The precisions of observables in the simplified electroweak fit where we neglect non-oblique corrections and parametrize the new physics contributions to electroweak observables in $S$ and $T$. The first five observables in the table and $S, T$ are free in the fit while the remaining three are determined by the free ones. We quote the precisions of current, high luminosity LHC and ILC measurements as well as the current central values. Entries that do not display a theory uncertainty either incorporate it into the experimental error bar or have a small enough theoretical uncertainty that it can be neglected. At the ILC, the non-negligible theory uncertainties of the derived observables $m_W, \sin^2 \theta_{\rm eft}^\ell$ and $\Gamma_Z$ come from unknown four-loop contributions assuming that in the future, the electroweak three-loop correction will be computed. In Sec.~\ref{sec:details}, we will explain in details the origins of all the numbers we used. }  
\label{tab:observables}
\end{table}

\begin{table}[!h]\footnotesize
\centering
\setlength{\tabcolsep}{.3em}
\begin{tabular}{|c|c|c|c|}
\hline
& TLEP-$Z$ & TLEP-$W$ & TLEP-$t$\\
\hline
$\alpha_s(M_Z^2)$ &
  $\pm 1.0 \times 10^{-4}$~\cite{Lepage:2014fla} & 
 $\pm 1.0 \times 10^{-4}$~\cite{Lepage:2014fla} &
 $\pm 1.0 \times 10^{-4}$~\cite{Lepage:2014fla} 
\\
$\Delta\alpha_{\rm had}^{(5)}(M_Z^2) $ &
$\pm 4.7 \times 10^{-5}$ & 
  $\pm 4.7 \times 10^{-5}$&
  $\pm 4.7 \times 10^{-5}$ 
\\
$m_Z$ [GeV] &
  $\pm 0.0001_{\rm exp}$~\cite{Gomez-Ceballos:2013zzn} & 
  $\pm 0.0001_{\rm exp}$~\cite{Gomez-Ceballos:2013zzn} &
   $\pm 0.0001_{\rm exp}$~\cite{Gomez-Ceballos:2013zzn}
\\
$m_t$ [GeV] (pole)&
 $\pm 0.6_{\rm exp} \pm 0.25_{\rm th}$~\cite{Baak:2014ora} & 
 $\pm 0.6_{\rm exp} \pm 0.25_{\rm th}$~\cite{Baak:2014ora} &
   $\pm 0.02_{\rm exp} \pm 0.1_{\rm th}$~\cite{Gomez-Ceballos:2013zzn,Baak:2014ora} 
\\
$m_h$ [GeV] &
     $<\pm 0.1$& 
   $<\pm 0.1$&
   $<\pm 0.1$
\\
\hline
$m_W$ [GeV] &
      $\left(\pm 8_{\rm exp} \pm 1_{\rm th}\right) \times 10^{-3}$~\cite{Baak:2014ora,Freitas:2013xga}& 
  $\left(\pm 1.2_{\rm exp} \pm 1_{\rm th}\right) \times 10^{-3}$~\cite{Baak:2013fwa,Freitas:2013xga} &
   $\left(\pm 1.2_{\rm exp} \pm 1_{\rm th}\right) \times 10^{-3}$~\cite{Baak:2013fwa,Freitas:2013xga} 
   \\
   $\sin^2\theta^{\ell}_{\rm eff}$  &
     $\left(\pm 0.3_{\rm exp} \pm 1.5_{\rm th}\right) \times 10^{-5}$~\cite{Baak:2013fwa,Freitas:2013xga} & 
     $\left(\pm 0.3_{\rm exp} \pm 1.5_{\rm th}\right) \times 10^{-5}$~\cite{Baak:2013fwa,Freitas:2013xga} & 
     $\left(\pm 0.3_{\rm exp} \pm 1.5_{\rm th}\right) \times 10^{-5}$~\cite{Baak:2013fwa,Freitas:2013xga} 
   \\
   $\Gamma_{Z}$ [GeV] &
     $\left(\pm 1_{\rm exp} \pm 0.8_{\rm th}\right) \times 10^{-4}$~\cite{Gomez-Ceballos:2013zzn,Freitas:2014hra} & 
  $\left(\pm 1_{\rm exp} \pm 0.8_{\rm th}\right) \times 10^{-4}$~\cite{Gomez-Ceballos:2013zzn,Freitas:2014hra} &
   $\left(\pm 1_{\rm exp} \pm 0.8_{\rm th}\right) \times 10^{-4}$~\cite{Gomez-Ceballos:2013zzn,Freitas:2014hra} 
   \\
\hline
\end{tabular}
\caption{The precisions of electroweak observables in the simplified electroweak fit at TLEP. We consider three scenarios: TLEP-$Z$: $Z$ pole measurement (including measurements with polarized beams); TLEP-$W$: $Z$ pole measurement plus scan of $WW$ threshold; TLEP-$t$: $Z$ pole measurement, $W$ threshold scan and top threshold scan. The TLEP experimental precisions are taken from either~\cite{Gomez-Ceballos:2013zzn} and~\cite{Baak:2013fwa}, where we always chose the more conservative numbers. Entries that do not display a theory uncertainty either incorporate it into the experimental uncertainty or have a small enough theoretical uncertainty that it can be neglected. Theoretical uncertainties may matter for $m_Z$ at TLEP, but we lack a detailed estimate and have not incorporated them. Similar to ILC, the non-negligible theory uncertainties of the derived observables $m_W, \sin^2 \theta_{\rm eft}^\ell$ and $\Gamma_Z$ come from unknown four-loop contributions assuming that in the future, the electroweak three-loop correction will be computed. In Sec.~\ref{sec:details}, we will explain in details the origins of all the numbers we used.}  
\label{tab:observables2}
\end{table}

\section{Prospects for CEPC Electroweak Precision}
\label{sec:cepc}
In this section, we will study the prospects of electroweak precision measurements at the Circular Electron Positron Collider (CEPC). So far there is very limited study of CEPC in the literature. We will present the first estimate of the reach for new physics of the electroweak program at CEPC based on the talk in~\cite{LiangTalk}. The precisions of the electroweak observables used in the simplified fit are summarized in Table.~\ref{tab:cepc}.\footnote{The summary table in the talk~\cite{LiangTalk} quotes an achievable precision for $\sin^2 \theta_{\rm eff}^\ell$ of 0.01\%, but based on the earlier slides and personal communication with Zhijun Liang we expect that 0.02\% is a reasonably optimistic choice.}
The $W$ mass precision is based on the direct measurement in $\sqrt{s} = 240$ GeV running with 100 fb$^{-1}$ integrated luminosity. The precisions of $Z$ mass and weak mixing angle are estimated for an energy scan on and around the $Z$ pole with $(100 - 1000)$ fb$^{-1}$ luminosity on the $Z$ pole and 10 fb$^{-1}$ for 6 energy points close to the $Z$ pole. The weak mixing angle is derived from the forward-backward asymmetry $A_{FB}$ of the $b$ quark, which is determined from fits to the differential cross-section distribution $d \sigma/d \cos \theta \propto 1 + \cos2 \theta + 8/3 A_{FB} \cos \theta$.
We will also present estimates of Higgs couplings precisions in Table~\ref{tab:CEPChiggs} of Section~\ref{sec:CEPChiggs}.

\begin{table}[!h]
\centering
\setlength{\tabcolsep}{.3em}
\begin{tabular}{|c|c|}
\hline
& CEPC\\
\hline
$\alpha_s(M_Z^2)$ &
  $\pm 1.0 \times 10^{-4}$~\cite{Lepage:2014fla} \\
$\Delta\alpha_{\rm had}^{(5)}(M_Z^2) $ &$\pm 4.7 \times 10^{-5}$ \\
$m_Z$ [GeV] &$\pm (0.0005 - 0.001)$~\cite{LiangTalk}  \\
$m_t$ [GeV] (pole)&$\pm 0.6_{\rm exp} \pm 0.25_{\rm th}$~\cite{Baak:2014ora} \\
$m_h$ [GeV] & $<\pm 0.1$ \\
\hline
$m_W$ [GeV] & $\left(\pm (3-5)_{\rm exp} \pm 1_{\rm th}\right) \times 10^{-3}$~\cite{LiangTalk,Awramik:2003rn,Freitas:2013xga} \\
$\sin^2\theta^{\ell}_{\rm eff}$  &$\left(\pm (4.6-5.1)_{\rm exp} \pm 1.5_{\rm th}\right) \times 10^{-5}$~\cite{LiangTalk,Awramik:2006uz,Freitas:2013xga}  \\
$\Gamma_{Z}$ [GeV] & $\left(\pm (5 - 10)_{\rm exp} \pm 0.8_{\rm th}\right) \times 10^{-4}$~\cite{LiangTalk,Freitas:2014hra} \\
\hline
\end{tabular}
\caption{The precisions of electroweak observables in the simplified electroweak fit at CEPC. The experimental uncertainties are mostly taken from~\cite{LiangTalk}. Entries that do not display a theory uncertainty either incorporate it into the experimental error bar or have a small enough theoretical uncertainty that it can be neglected. Similar to ILC and TLEP, the non-negligible theory uncertainties of the derived observables $m_W, \sin^2 \theta_{\rm eft}^\ell$ and $\Gamma_Z$ come from unknown four-loop contributions assuming that in the future, the electroweak three-loop correction will be computed. For $\Gamma_Z$, we assumed that it has the same experimental uncertainty as $m_Z$. }  
\label{tab:cepc}
\end{table}

\begin{figure}[!h]\begin{center}
\includegraphics[width=0.4\textwidth]{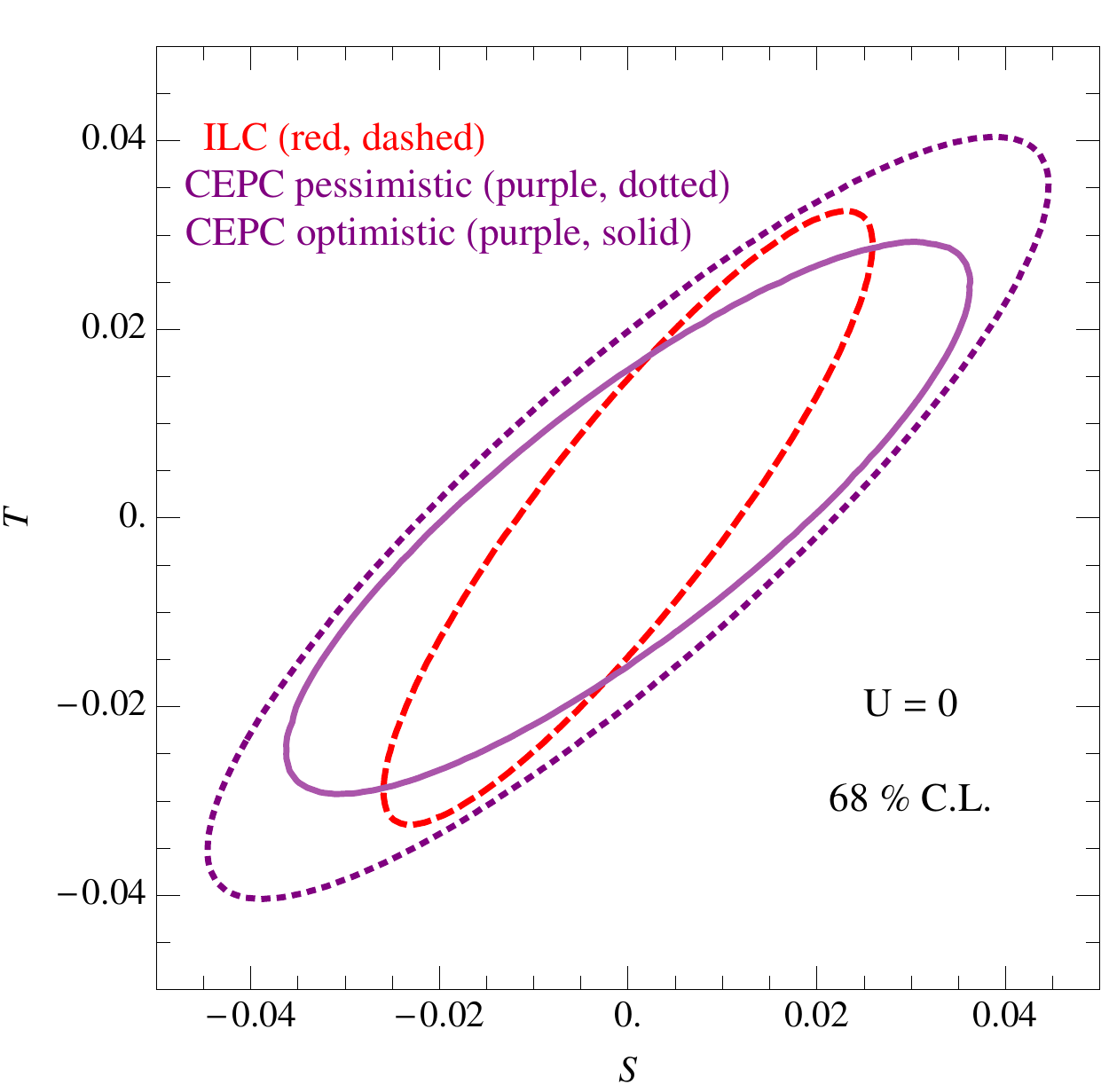} 
\end{center}
\caption{68\% C.L. contours of $S$ and $T$ for CEPC using the simplified fit with inputs in Table \ref{tab:cepc}. For comparison, we also show the ILC allowed region (red dashed line) derived in Sec~\ref{sec:oblique}.We set the best fit point to be $S=T=0$, which corresponds to the current SM values. The dotted purple contour is derived with the numbers at the higher ends of the estimated ranges in Table \ref{tab:cepc} while the solid purple contour is derived with those at the lower ends. }
\label{fig:cepc}
\end{figure}%

We also performed a profile likelihood fit and present the allowed $(S, T)$ region for CEPC at 68\% C.L. in Fig.~\ref{fig:cepc}. For comparison, we put the ILC result in the same plot. For the more optimistic evaluation in which all precisions take the lower end values of the estimated ranges in Table~\ref{tab:cepc}, the ILC and CEPC have similar sensitivities to new physics. For the more pessimistic evaluation based on precisions at the higher ends of the estimated ranges, the CEPC allows larger $S$ mostly because of the worse precision of $\sin^2\theta^{\ell}_{\rm eff}$ compared to ILC.

\subsection{Hypothetical Improvements of CEPC EWPT}
\label{subsec:cepc_imp}

In this section, we will consider possible improvements of electroweak observable precisions at CEPC and study how they affect the CEPC's sensitivity to new physics. There are four potential improvements of electroweak observables: $m_t$, $m_W$, $\sin^2\theta_{\rm eff}^\ell$ and $\Gamma_Z$ (together with $m_Z$), which are listed in Table~\ref{tab:cepcimp}. 

The top quark mass gives the largest parametric uncertainties on the derived SM observables in the global fit (more details could be found in Sec.~\ref{sec:par}) and thus improving its precision might improve the fit. In the fit for CEPC above, we assumed the precision of the top mass after the HL-LHC running. A top threshold scan is not included in the current CEPC plan, so CEPC itself cannot improve the precision of $m_t$. However, a top threshold scan is part of the ILC plan. The possibility exists if the ILC program with the top threshold scan is implemented before or at the same time of CEPC, the input value of $m_t$ precision for the CEPC electroweak fit could be improved by a factor of $\sim 10$. The precision of the $W$ mass could be slightly improved by a $WW$ threshold scan to 2 MeV~\cite{LiangTalk}. Finally, the uncertainty of $\sin^2\theta_{\rm eff}^\ell$ in the current CEPC plan is still dominated by the statistical uncertainty, which is $0.02\%$ while the systematic uncertainty is $0.01\%$. If the luminosity of the off-peak $Z$ running could be increased by a factor 4 to 40 fb$^{-1}$ (at each energy), the overall uncertainty of $\sin^2\theta_{\rm eff}^\ell$ could be reduced down to $0.01\%$, which is $2.3 \times 10^{-5}$. Another possible way to reduce the uncertainty of $\sin^2\theta_{\rm eff}^\ell$ down to $0.01\%$ is to use polarized electron/positron beams, which would require more infrastructure. If CEPC could perform energy calibration using the resonant spin depolarization method, which will be described in Sec.~\ref{sec:zmass}, at the collision time as in the TLEP plan, the systematic uncertainties of $\Gamma_Z$ and $m_Z$ could potentially be reduced as low as 100 keV. 

\begin{figure}[!h]\begin{center}
\includegraphics[width=0.3\textwidth]{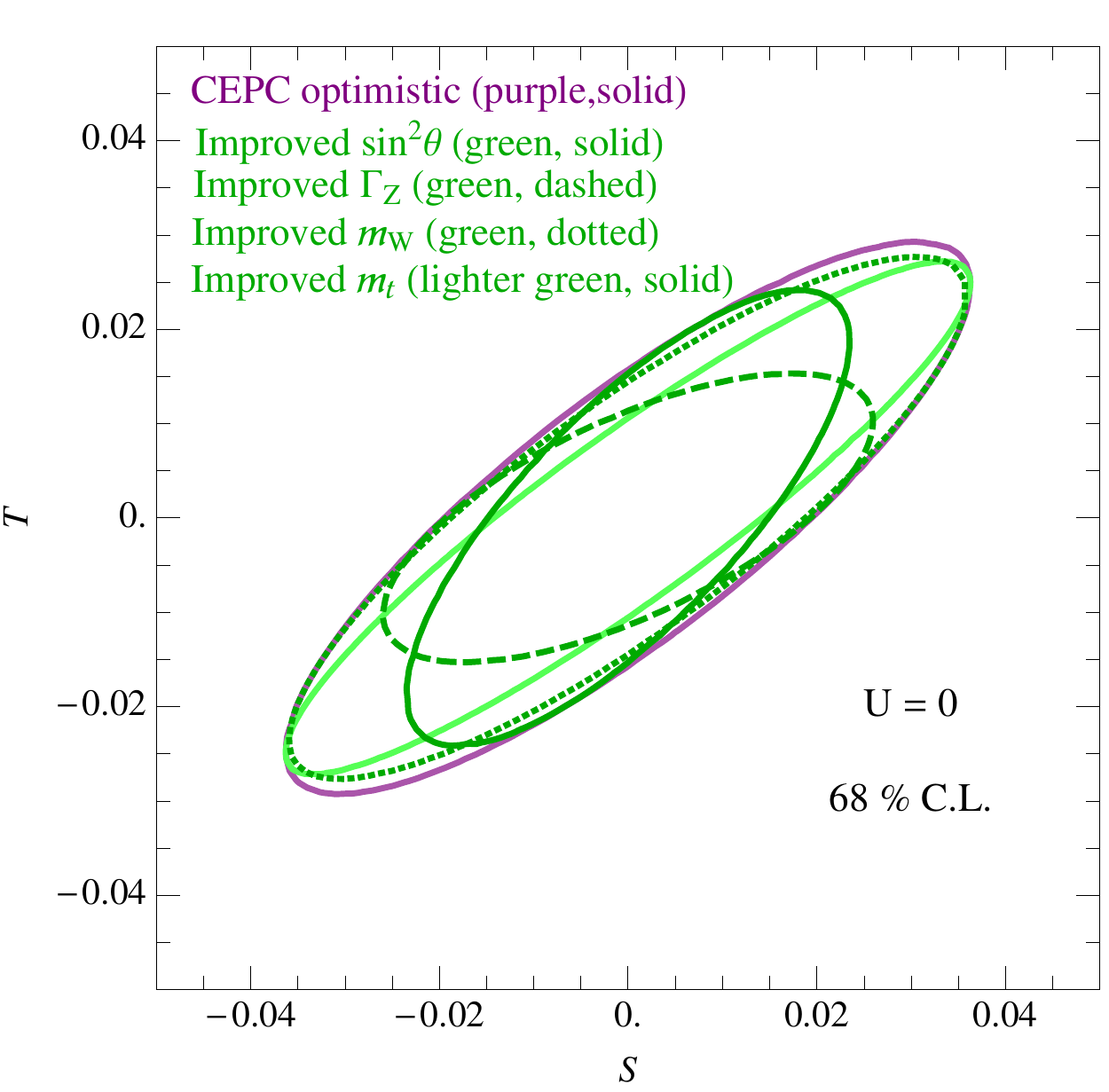} \quad \includegraphics[width=0.3\textwidth]{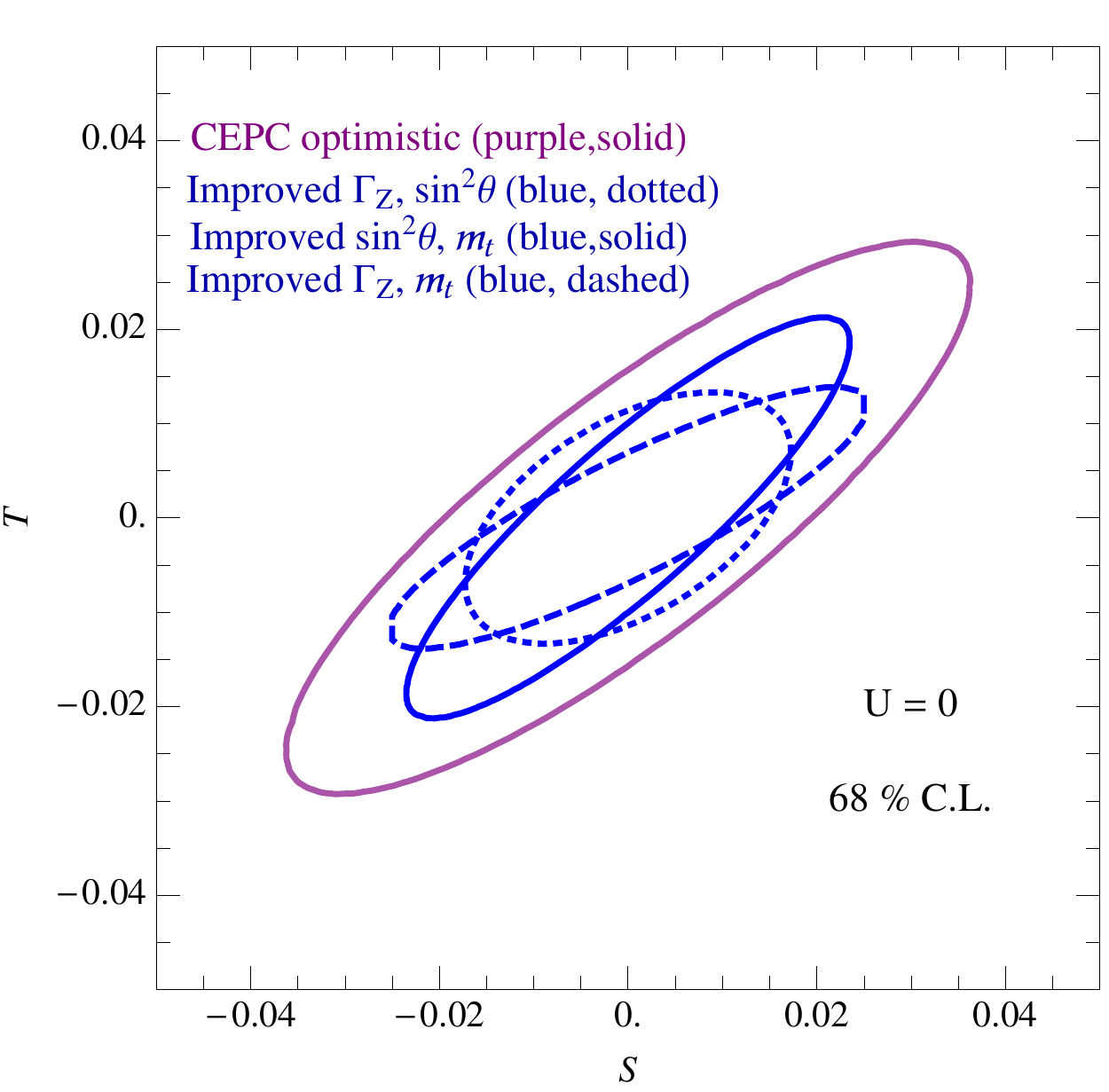} \quad \includegraphics[width=0.3\textwidth]{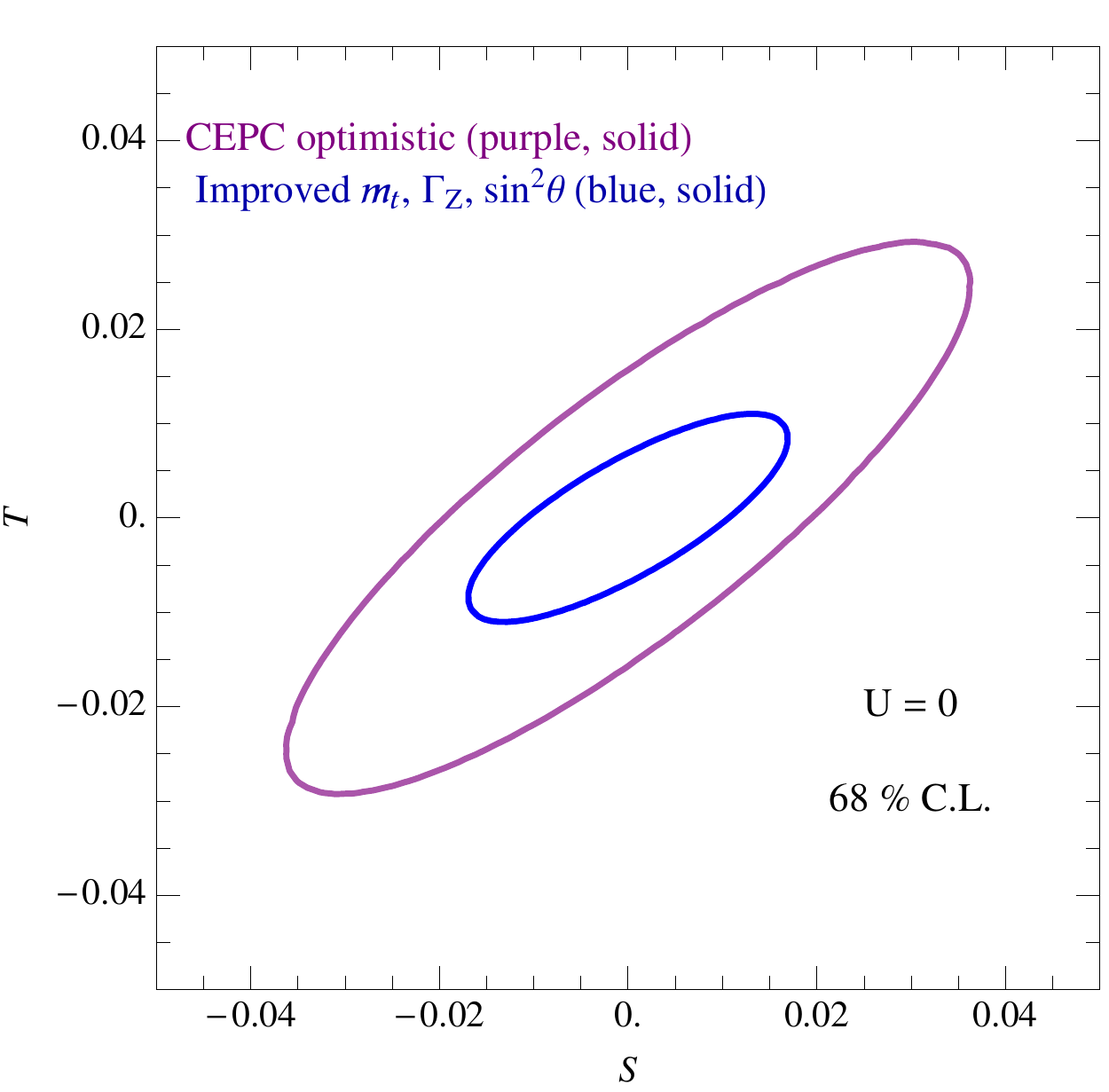}
\end{center}
\caption{68\% C.L. contours of $S$ and $T$ for CEPC with one of the four parameters $m_t, m_W, \sin^2 \theta_W,$ or  $\Gamma_Z$ improved (left), two improved (middle), and three of them improved (right) relative to the optimistic case of Fig.~\ref{fig:cepc}. The improved values are listed in Table \ref{tab:cepcimp}. One could see from the left panel that improving $m_W$ only does not help improve the sensitivity. In the middle and right panels, we don't show ellipses with improved $m_W$ together with other improved observables because improved $m_W$ precision does not help much on top of the improvements due to the other improved observables.  
For comparison, we also showed in each plot 68\% C.L. contours of $S$ and $T$ for CEPC with the most optimistic inputs in Table \ref{tab:cepc}.}
\label{fig:cepcimp}
\end{figure}

\begin{figure}[!h]\begin{center}
\includegraphics[width=0.5\textwidth]{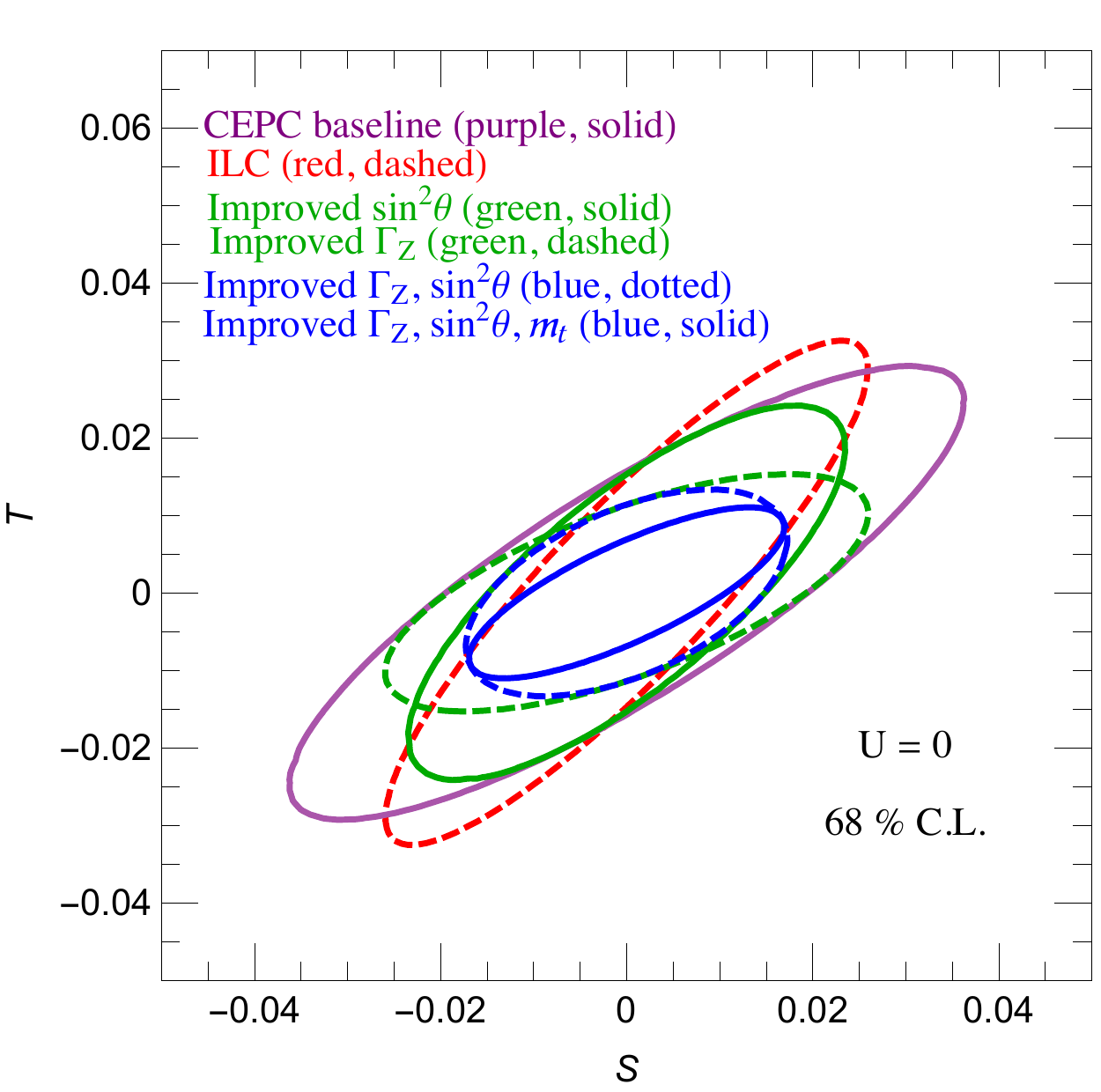}
\end{center}
\caption{68\% C.L. contours of $S$ and $T$ for ILC (red dashed), the optimistic case of current CEPC plan (named as the CEPC baseline in the figure; purple solid), the optimistic CEPC plan with $\sin^2 \theta_W$ (green solid) or $\Gamma_Z$ (green dashed) improved, both $\sin^2 \theta_W$ and $\Gamma_Z$ improved (blue dotted), and three observables $\sin^2 \theta_W$, $\Gamma_Z$ and $m_t$ improved (blue solid).}
\label{fig:cepcmoney}
\end{figure}

Now we want to assess how these potential improvements affect the CEPC's sensitivity and whether it is worthwhile to implement them. We performed fits with one, two or three of the improvements in precision discussed above, always relative to the optimistic case from Table~\ref{tab:cepc}. The results are shown in Fig.~\ref{fig:cepcimp}. From the figure, one could see that the improvement of $m_W$ precision alone does not help.  Each of the other three improvements could constrain $S$ or $T$ a bit more. Combining improvements in the $\Gamma_Z$ and $\sin^2\theta_{\rm eff}^\ell$ precisions lead to an increase in the sensitivity to $S$ and $T$ by a factor of 2.  Further combination with a improved measurement of $m_t$ leads to a small improvement in the constraint. We summarize the potential major improvements of sensitivities in the $S$ and $T$ plane in Fig.~\ref{fig:cepcmoney}. The improved CEPC measurements could outperform the ILC ones in the $S$ and $T$ reach because of a better determination of $m_Z$ and $\Gamma_Z$ from a better energy calibration. As will be explained in the next section, a circular collider could do a better job of energy calibration due to the resonant spin depolarization technique.

\begin{table}[!h]\small
\centering
\setlength{\tabcolsep}{.3em}
\begin{tabular}{|c|c|c|c|c|}
\hline
CEPC & $m_t$ [GeV] & $m_W$ [GeV] & $\sin^2\theta_{\rm eff}^\ell$ & $\Gamma_Z$ [GeV]\\
\hline
Improved Error & $\pm 0.03_{\rm exp} \pm 0.1_{\rm th}$ & $(\pm2_{\rm exp} \pm 1_{\rm th}) \times 10^{-3}$ & $(\pm2.3_{\rm exp} \pm 1.5_{\rm th})  \times 10^{-5}$ & $(\pm 1_{\rm exp} \pm 0.8_{\rm th}) \times 10^{-4}$  \\
\hline
\end{tabular}
\caption{Hypothetical improvements of electroweak observable precisions for CEPC. The improvement of $m_t$ precision could come from the ILC top threshold scan if it happened before or at the same time as CEPC; $m_W$ precision could be improved slightly by a $WW$ threshold scan~\cite{LiangTalk}; $\sin^2\theta_{\rm eff}^\ell$ precision could be improved if the statistical uncertainty is reduced to be smaller than the systematic uncertainty, which is $0.01\%$~\cite{LiangTalk}. $\Gamma_Z (m_Z)$ precision could be improved if the systematic uncertainty from the energy calibration could be reduced down to the TLEP projection. }  
\label{tab:cepcimp}
\end{table}

\section{Details of Electroweak Fit}
\label{sec:details}

In this section we will explain the details of a number of uncertainties that have gone into the fit in Sec.~\ref{sec:oblique}. 

\subsection{Nuisance Parameters}
\label{sec:nuisance}

\subsubsection{The Top Mass $m_t$}

Recently, the first combination of Tevatron and LHC top mass measurements reported a result of $173.34 \pm 0.76$ GeV, with the error bar combining statistical and systematic uncertainties~\cite{ATLAS:2014wva}. New results continue to appear, with a recent CMS combination reporting $172.38\pm0.10~{\rm (stat.)}\pm0.65~{\rm (syst.)}$ GeV~\cite{CMS:2014hta} and a D0 analysis finding $174.98 \pm 0.76$ GeV~\cite{Abazov:2014dpa}. These results have similar error bars but fairly different central values, which may be a statistical fluke or may in part reflect ambiguities in defining what we mean by the top mass (see~\cite{Hoang:2008xm} and Appendix C of~\cite{Buckley:2011ms}). This suggests that we proceed with some caution in assigning an uncertainty to the top mass in any precision fit.

The relevant physics issues have been reviewed recently in refs.~\cite{Juste:2013dsa,Agashe:2013hma,Moch:2014tta}. At the LHC, kinematic measurements are expected to reach a precision of 0.5 or 0.6 GeV on the top mass, but theoretical uncertainty remains in understanding how the measured mass relates to well-defined schemes like the $\overline{\rm MS}$ mass. Other observables like the total cross section are easier to relate to a choice of perturbative scheme, but will have larger uncertainties. The top mass is a very active area of research, in part for its importance in questions of vacuum stability in the Standard Model (see, for example, refs.~\cite{Bezrukov:2012sa,Degrassi:2012ry,Buttazzo:2013uya,Andreassen:2014gha}). As a result, we can expect continued progress in understanding how to make the best use of the LHC's large sample of top quark data to produce more accurate mass determinations. For a sampling of recent ideas in this direction, see~\cite{ATLASMinSukKimforthe:2014hba,Kawabataa:2014osa,Frixione:2014ala, Argyropoulos:2014zoa}. We will follow ref.~\cite{Baak:2014ora} in assuming that the LHC will achieve a measured precision of 0.6 GeV and that further experimental and theoretical effort will reduce the theoretical uncertainty on the meaning of this number to 0.25 GeV. We will also use their estimate of the current theoretical uncertainty as 0.5 GeV, although we suspect this is overly optimistic.

At a linear collider, a threshold scan may be used to simultaneously fit the top mass and width, $\alpha_s$, and the top Yukawa coupling. Recent estimates include refs.~\cite{Seidel:2013sqa,Horiguchi:2013wra}. A statistical precision of about 30 MeV is widely agreed to be possible, but systematic uncertainties including the luminosity spectrum and beam energy add to this. The recent review article~\cite{Juste:2013dsa}, for instance, attributes a 50 MeV uncertainty from the luminosity spectrum, whereas ref.~\cite{Seidel:2013sqa} gives a preliminary estimate of 75 MeV for this uncertainty. Furthermore, converting from the 1S scheme to $\overline{\rm MS}$ scheme adds a theoretical uncertainty of about 100 MeV. For the ILC, we will again follow ref.~\cite{Baak:2014ora} by assigning an experimental uncertainty of 30 MeV and a theoretical uncertainty of 100 MeV for the ILC measurement, despite its optimism regarding experimental systematics. The TLEP report estimates that a 10 to 20 MeV experimental precision can be attained on the top quark mass~\cite{Gomez-Ceballos:2013zzn}. Again, the theory uncertainty is dominant. We choose to use the 20 MeV estimated precision but also include a 100 MeV theoretical uncertainty. We find that omitting this theory uncertainty does not dramatically change the reach, mainly due to the dominance of other systematic uncertainties such as $\Delta \alpha_{\rm had}$.

\subsubsection{The Hadronic Contribution $\Delta \alpha_{\rm had}$}

The fine structure constant $\alpha$ measured at low energies is an input to electroweak precision fits, but its value must be extrapolated to high energies. The main uncertainty in doing so is the hadronic contribution to the running, denoted $\dahad(m_Z^2)$ and defined via:
\beq
\alpha(q^2) = \frac{\alpha(0)}{1 - \dahad(q^2) - \Delta \alpha_{\rm lep}(q^2)}.
\eeq 
(The superscript refers to the five flavors of quark that contribute.) This quantity is of great interest not only for its role in electroweak precision fits, but also because of its close link to the hadronic vacuum polarization contributions to muon $g-2$, which play a key role in understanding the amount of tension between the measured value and the Standard Model prediction. Several recent determinations of $\dahad(m_Z^2)$ exist~\cite{Davier:2010nc,Hagiwara:2011af,Burkhardt:2011ur,Jegerlehner:2011mw,Bodenstein:2012pw}. The analogous leptonic contribution $\Delta \alpha_{\rm lep}(m_Z^2)$ is known at 3 loops to be $314.97686 \times 10^{-4}$~\cite{Steinhauser:1998rq}.

Determinations of $\dahad(m_Z^2)$ typically rely on a mix of data-driven estimates and theoretical calculation to obtain the integrand of a dispersion relation for the running coupling in terms of the principal value of an integral~\cite{Cabibbo:1961sz}:
\beq
\dahad(q^2) = -\frac{\alpha q^2}{3\pi} {\rm P} \int^\infty_{4 m_\pi^2} \frac{R_{\rm had}(s') ds'}{s'(s' - q^2)},
\eeq 
where $R_{\rm had} = \sigma(e^+ e^- \to \gamma^* \to {\rm hadrons})/\sigma(e^+ e^- \to \gamma^* \to \mu^+ \mu^-)$. Notice that this integral involves the cross section at physical (timelike) momenta. The integral is generally broken into pieces: at large $s'$, where hadronic resonances are well-approximated by a partonic continuum, perturbative QCD can be used. At small $s'$, hadronic resonances like the $\rho$ meson are important, and $R_{\rm had}$ is usually taken from data. Alternatively, one can make use of the Adler function, which is the analytic continuation of the dispersion integral above to Euclidean momenta $Q^2 = -q^2 > 0$. At large $Q^2$ this function can be computed from perturbation theory. At small $Q^2$, future lattice studies may determine this function with sufficient accuracy to allow a precise computation of $\dahad(m_Z^2)$ independent of experimental data on the cross section~\cite{Bodenstein:2012pw}. For now, however, experimental measurements are a major input and the major source of uncertainty.

The total cross section $\sigma(e^+ e^- \to \gamma^* \to {\rm hadrons})$, as a function of center-of-mass energy, has been measured both by scanning the center-of-mass energy of the collider itself and by radiative return, i.e. studying $e^+ e^- \to \gamma \gamma^*$ as a function of the on-shell ISR photon's energy (or, equivalently, virtuality of the off-shell $\gamma$)~\cite{Binner:1999bt}. The latter technique allows modern colliders like KLOE~\cite{Ambrosino:2010bv} and BaBar~\cite{Aubert:2009ad} that operate at fixed center-of-mass energy to probe the cross section at lower energies. It is somewhat less clean (suffering from, for instance, the problem of separating FSR from ISR photons), but allows the use of very large data sets from recent fixed-energy high-luminosity experiments. Various groups have combined such data in fits with data from experiments like CMD-2~\cite{Akhmetshin:2003zn} and BES~\cite{Bai:1999pk,Bai:2001ct} that scan in energy.

Among the recent determinations, the highest accuracy is claimed by ref.~\cite{Bodenstein:2012pw}, which uses new perturbative calculations of heavy-quark contributions and quotes
\beq
\dahad(m_Z^2) = \left(276.5 \pm 0.8\right) \times 10^{-4}.
\eeq
The largest error bar determined recently is from ref.~\cite{Burkhardt:2011ur}, which quotes $\dahad(m_Z^2) = \left(275.0 \pm 3.3\right) \times 10^{-4}$. Their analysis makes use of BaBar data only in the region around the $\rho$ peak and not at higher energies, where many different exclusive final states open up. The numbers quoted in ref.~\cite{Davier:2010nc,Hagiwara:2011af,Jegerlehner:2011mw} agree well with ref.~\cite{Bodenstein:2012pw} in central value and have somewhat bigger error bars ($1.0~{\rm to}~1.4 \times 10^{-4}$). Many other determinations of $\dahad(m_Z^2)$ are tabulated in the PDG review~\cite{pdgreview}. Recent studies of electroweak precision at future colliders have assumed that the error bar on $\dahad(m_Z^2)$ can be decreased to $5.0 \times 10^{-5}$~\cite{Baak:2013fwa, MishimaTalk}. This seems very reasonable, given the steady progress so far and the possibility for additional input data from $e^+ e^-$ colliders operating at $\sqrt{s}$ below 10 GeV to improve on the current result. For example, data from VEPP-2000 and BESIII are expected to reduce the uncertainty on the hadronic cross section below 2 GeV by a factor of between 2 and 3~\cite{Blum:2013xva}. We will follow the other recent studies in projecting a future uncertainty of $5.0 \times 10^{-5}$, but suspect that it may even prove to be overly conservative.

\subsubsection{The Strong Coupling $\alpha_s$ and the Charm and Bottom Masses}

The value of the strong coupling constant $\alpha_s$ is one of the major sources of uncertainty in precision tests of Higgs boson properties. The current status of $\alpha_s$ measurements was recently reviewed in refs.~\cite{Pich:2013sqa,Moch:2014tta}. The Particle Data Group's current world average is~\cite{Beringer:1900zz}
\beq
\alpha_s(m_Z^2) = 0.1185 \pm 0.0006,
\eeq
whereas ref.~\cite{Pich:2013sqa} quotes $0.1186 \pm 0.0007$. There may be some lingering systematic issues in the data, with DIS determinations being characteristically low, but the fit is relatively insensitive to dropping DIS. 

Prospects for future improvements using the lattice were reviewed in ref.~\cite{Lepage:2014fla}. We will follow it in taking the currently measured charm and bottom quark masses from the lattice result~\cite{McNeile:2010ji}:
\beq
m_b(10~{\rm GeV}, n_f = 5) & = & 3.617 \pm 0.025~{\rm GeV}, \\
m_c(3~{\rm GeV}, n_f = 4) & = & 0.986 \pm 0.006~{\rm GeV}.
\eeq
According to ref.~\cite{Lepage:2014fla}, feasible improvements estimated from a combination of perturbative calculations, decrease in lattice spacing, and increased lattice statistics reduce the error bars to $\delta \alpha_s(m_Z^2) \approx 9 \times 10^{-5}$, $\delta m_b(10~{\rm GeV}) \approx 0.003~{\rm GeV}$, and $\delta m_c(3~{\rm GeV}) \approx 0.002~{\rm GeV}$. Furthermore, TLEP hopes to directly measure $\alpha_s(m_Z)$ at $10^{-4}$ accuracy~\cite{Gomez-Ceballos:2013zzn}. In this case, the theoretical accuracy of SM predictions for Higgs properties can be reduced below the measurement accuracies attained at the ILC or TLEP.

\subsubsection{The $Z$ and Higgs Masses}
\label{sec:zmass}

For the Higgs mass, we follow ref.~\cite{Baak:2014ora} in averaging recent ATLAS~\cite{Aad:2014aba} and CMS~\cite{CMS:2014ega} results to obtain
\beq
m_h = 125.14 \pm 0.24~{\rm GeV}.
\eeq
We further follow ref.~\cite{Baak:2014ora} in assuming an eventual uncertainty of 0.1 GeV or below in the LHC's Higgs mass measurement. (The precise error bar makes little difference in the fit.)

The best current measurement of the $Z$ mass is $m_Z = 91.1875 \pm 0.0021$ GeV from LEP~\cite{ALEPH:2005ab}. The statistical error is about 1.2 MeV while the dominating systematics uncertainty comes from the energy calibration. At circular colliders such as LEP and TLEP, the precise determination of the beam energy is based on the technique of resonant spin depolarization~\cite{ALEPH:2005ab}. As charged particles move in the magnetic field that bends them around the circular tunnel, the average spin of the polarized bunches precesses. The beam energy is proportional to the number of times the spins precess per turn. Then one could observe a depolarization which occurs when a weak oscillating radial magnetic field is applied to the spins, achieving a resonance that allows an accurate measurement of the spin precession frequency. The intrinsic uncertainty of this method is about 100 keV on the beam energy at the time of the measurement. However, at LEP, the calibration was performed outside the collision period and then extrapolated back to the collision time. During the period of calibration, the movement of LEP equipment due to tidal effects, water level in Lake Geneva, and even rainfall in the nearby mountains inflated the error bar of $m_Z$ to 1.7 MeV and that of $\Gamma_Z$ to 1.2 MeV~\cite{ALEPH:2005ab}. The remaining errors are theoretical uncertainties including initial state radiation, fermion-pair radiation and line-shape parametrization, which add up to about 400 keV~\cite{ALEPH:2005ab}. 

At TLEP, the energy calibration uncertainty could be reduced to 100 keV because it is possible to calibrate at the time of collision. The number of bunches is large enough that one could apply resonant depolarization to a few bunches---say 100 bunches---which would not collide, while the other bunches are colliding. The systematic uncertainty related to the extrapolation at LEP before would then disappear. Currently the largest theory uncertainty of order a few hundred keV arises from corrections of leptonic pair radiation of order ${\cal O} (\alpha^3)$ and higher as well as an approximate treatment of hadronic pair radiation~\cite{Arbuzov:1999uq,Arbuzov:2001rt}. Certainly the computations need to be improved by at least a factor of about 5 before the next generation circular $e^+e^-$ collider is built to bring the total uncertainty down to 100 keV as expected in the TLEP report~\cite{Gomez-Ceballos:2013zzn}.

At the ILC, however, the energy calibration is completely different because there is no resonant spin depolarization in a linear collider! A magnetic spectrometer could measure the beam energy with resolution of a few $10^{-4}$ and M{\o}ller scattering method could measure $\Delta E/E \approx 10^{-5}$ in the vicinity of the $Z$ peak whose position is cross-calibrated using the LEP measured $Z$ mass~\cite{AguilarSaavedra:2001rg}. Thus at ILC, the precision of $m_Z$ will not be improved while that of $\Gamma_Z$ could be improved by a factor around 2. 

\subsection{Non-nuisance Parameters}
\label{sec:nonnuisance}
 In this section, we will review current and future experimental and theoretical uncertainties of the three derived SM observables used in our simplified fit: $m_W$, $\sin^2\theta_{\rm eff}^\ell$ and $\Gamma_Z$. The experimental uncertainties have already been collected with details in a Snowmass paper~\cite{Baak:2013fwa}. Here we just offer a quick review for the completeness of our discussions. 

\subsubsection{Experimental Uncertainties}
\label{sec:exp}
The average measured $W$ boson mass is $80.385 \pm 0.015$ GeV by the LEP and Tevatron experiments~\cite{Beringer:1900zz}. At ILC, there are three options to measure the $W$ mass more precisely: polarized threshold scan of the $W^+W^-$ cross section, kinematically-constrained reconstruction of $W^+W^-$ and direct measurement of the hadronic mass in full hadronic or semi-leptonic $W^+W^-$ events. The target uncertainties for each method could be found in~\cite{Baak:2013fwa}. An overall 5 MeV experimental uncertainty is perceived to be possible. At TLEP, given the potential big reduction in the energy calibration uncertainty as explained in Sec.~\ref{sec:zmass}, $m_W$'s uncertainty is statistics dominated. With a thorough scan at the $W^+W^-$ threshold, a 1 MeV uncertainty is supposed to be achievable at TLEP per experiment and 500 keV from a combination of four experiments. In our fit, we took the more conservative number 1 MeV for TLEP. 

The current value of the weak mixing angle $\sin^2 \theta_{\rm eff}^\ell = (23153 \pm 16) \times 10^{-5}$ is derived from a variety of measurements at LEP and SLD. LEP measured leptonic and hadronic forward-backward asymmetries from a line-shape scan without longitudinally polarized beams. On the other hand, SLD could produce longitudinally polarized electron and unpolarized positron beams and measure the left-right beam polarization asymmetry directly. The measurements with the smallest uncertainties are the SLD measurement and the forward-backward asymmetry of $b$ quarks at LEP (which, however, are not in good agreement with each other). For both measurements, the statistical and systematic errors are of the same order, with the statistical error bar dominating. At both ILC and TLEP, it is expected that the Blondel scheme could facilitate a significantly more precise measurement of asymmetries without requiring an absolute polarization measurement~\cite{Blondel:1987wr}. What is needed is a precise determination of the polarization difference between the two beam helicity states. If the scheme is implemented, at both ILC and TLEP, the statistical errors will become subdominant and the systematic errors could be reduced to 0.006\% and 0.001\% respectively. 

The $Z$ width measured at LEP is $2.4952 \pm 0.0023$ GeV. The statistical error is about 2 MeV while the systematic error from energy calibration is 1.2 MeV. At ILC, as already discussed in Sec.~\ref{sec:zmass}, the relative precision of the beam spectrometer could reduce the error bar of $\Gamma_Z$ by a factor of 2 while the position of the $Z$ peak is calibrated using the LEP result. At TLEP, the statistical error is negligible while the systematic uncertainty could be reduced to 100 keV in principle due to a potentially much more precise energy calibration as discussed in Sec.~\ref{sec:zmass}.

\subsubsection{Parametric Uncertainties}
\label{sec:par}
Now we go through the theory uncertainties of $m_W$, $\sin^2\theta_{\rm eff}^\ell$ and $\Gamma_Z$: 

\begin{itemize}
{\item The presently most accurate prediction for $m_W$ is obtained by combining the complete two-loop result with the known higher-order QCD and electroweak corrections~\cite{Awramik:2003rn}. The remaining theory uncertainties are from higher-order corrections at order ${\cal O}(\alpha^2 \alpha_s), {\cal O} (\alpha^3)$ and ${\cal O} (\alpha \alpha_s^3)$ beyond the leading term in an expansion for asymptotically large values of $m_t$. This is estimated to be about 4 MeV~\cite{Awramik:2003rn}. With a full three-loop calculation including terms of order ${\cal O}(\alpha^2 \alpha_s)$ and ${\cal O} (\alpha^3)$, the theory error could be reduced to $\lesssim 1$ MeV, mainly from the four-loop QCD correction~\cite{Freitas:2013xga}. }
{\item A parametrization of $\sin^2\theta_{\rm eff}^\ell$ based on complete electroweak two-loop result is available as well~\cite{Awramik:2006uz}. Again the most relevant missing corrections are of the order ${\cal O}(\alpha^2 \alpha_s), {\cal O} (\alpha^3)$ and ${\cal O} (\alpha \alpha_s^3)$ beyond the leading term in an expansion for asymptotically large values of $m_t$. The theory error is estimated to be about $(4.4 - 4.7) \times 10^{-5}$. Once the complete three-loop calculation is done, the theory error will be of order ${\cal O} (\alpha \alpha_s^3)$ beyond the leading term in the large $m_t$ expansion, which is about $1.5 \times 10^{-5}$~\cite{Freitas:2013xga}. }
{\item For $\Gamma_Z$, there is still one missing piece at the two-loop order, which is the bosonic EW corrections of order ${\cal O}(\alpha^2_{\rm bos})$. This type of correction originates from diagrams without closed fermion loops. Parametrization of $\Gamma_Z$ based on known two-loop result could be found in~\cite{Freitas:2014hra}. Theory uncertainties also receive corrections at order ${\cal O}(\alpha^2 \alpha_s), {\cal O}(\alpha \alpha_s^2), {\cal O} (\alpha^3)$ and ${\cal O} (\alpha \alpha_s^3)$ beyond the leading $m_t$ terms. The unknown final state QCD correction at order ${\cal O} (\alpha_s^5)$ will also contribute. The total theory error adds up to about 0.5 MeV. Once the bosonic two-loop and the complete three-loop results are known, the theory error will be reduced to about 0.08 MeV. Notice that similar to $m_Z$, $\Gamma_Z$ also has theoretical uncertainties from initial state radiation, fermion-pair radiation and line-shape parametrization, which we do not include under the assumption that they will be accurately computed in the future.}
\end{itemize}

In our fits, we assumed that by the time when future $e^+e^-$ colliders are built, complete three-loop electroweak corrections have been computed and the theory uncertainties originate from the four-loop and higher-order corrections. 

\begin{table}[h]
\setlength{\tabcolsep}{.3em}
\begin{tabular}{|c|c|c|c|c|c|}
\hline
Current &$m_t$ & $m_Z$ & $m_h$& $\alpha_s$ & $\Delta\alpha_{\rm had}^{(5)}(M_Z^2)$ \\
\hline
$\delta m_W$ [MeV] & 4.6 &  2.6 & 0.1 &  0.4 & 1.5 \\
\hline
$\delta\sin^2\theta_{\rm eff}^\ell (10^{-5})$& 2.4 &  1.5 & 0.1 &  0.2 & 2.8 \\
\hline
$\delta\Gamma_Z$ [MeV] & 0.2 &  0.2 & 0.004 &  0.30 & 0.08 \\
\hline
\end{tabular}
~
\begin{tabular}{|c|c|c|c|c|c|}
\hline
ILC &$m_t$ & $m_Z$ & $m_h$& $\alpha_s$ & $\Delta\alpha_{\rm had}^{(5)}(M_Z^2)$ \\
\hline
$\delta m_W$ [MeV] & 0.2 &  2.6 & 0.05 &  0.06 & 0.9 \\
\hline
$\delta\sin^2\theta_{\rm eff}^\ell (10^{-5})$& 0.09 &  1.5 & 0.04 &  0.03 & 1.6 \\
\hline
$\delta\Gamma_Z$ [MeV] & 0.007 &  0.2 & 0.002 &  0.05 & 0.04 \\
\hline
\end{tabular}
~
\begin{tabular}{|c|c|c|c|c|c|}
\hline
TLEP-$Z$($W$) &$m_t$ & $m_Z$ & $m_h$& $\alpha_s$ & $\Delta\alpha_{\rm had}^{(5)}(M_Z^2)$ \\
\hline
$\delta m_W$ [MeV] & 3.6 &  0.1 & 0.05 &  0.06 & 0.9 \\
\hline
$\delta\sin^2\theta_{\rm eff}^\ell (10^{-5})$& 1.9 &  0.07 & 0.04 &  0.03 & 1.6 \\
\hline
$\delta\Gamma_Z$ [MeV] & 0.1 &  0.01 & 0.002 &  0.05 & 0.04 \\
\hline
\end{tabular}
~
\begin{tabular}{|c|c|c|c|c|c|}
\hline
TLEP-$t$ &$m_t$ & $m_Z$ & $m_h$& $\alpha_s$ & $\Delta\alpha_{\rm had}^{(5)}(M_Z^2)$ \\
\hline
$\delta m_W$ [MeV] & 0.1 &  0.1 & 0.05 &  0.06 & 0.9 \\
\hline
$\delta\sin^2\theta_{\rm eff}^\ell (10^{-5})$& 0.06 &  0.07 & 0.04 &  0.03 & 1.6 \\
\hline
$\delta\Gamma_Z$ [MeV] & 0.004 &  0.01 & 0.002 &  0.05 & 0.04 \\
\hline
\end{tabular}
\quad \quad \quad \quad \, 
\begin{tabular}{|c|c|c|c|c|c|}
\hline
CEPC &$m_t$ & $m_Z$ & $m_h$& $\alpha_s$ & $\Delta\alpha_{\rm had}^{(5)}(M_Z^2)$ \\
\hline
$\delta m_W$ [MeV] & 3.6 &  0.6-1.3 & 0.05 &  0.06 & 0.9 \\
\hline
$\delta\sin^2\theta_{\rm eff}^\ell (10^{-5})$& 1.9 &  0.4-0.7 & 0.04 &  0.03 & 1.6 \\
\hline
$\delta\Gamma_Z$ [MeV] & 0.1 &  0.05-0.1 & 0.002 &  0.05 & 0.04 \\
\hline
\end{tabular}
\caption{Parametric errors from each free parameter in the fit for current, ILC, TLEP-$Z$ (TLEP-$W$), TLEP-$t$ and CEPC scenarios.  }  
\label{tab:errors}
\end{table}

We list the breakdown of parametric uncertainties for current and future experimental scenarios in Table~\ref{tab:errors}. It is clear that currently the top and $Z$ boson masses are the dominant contributions to the parametric uncertainties. ILC can measure $m_t$ precisely, and $Z$ mass remains as the dominant uncertainty. When both are measured very precisely at TLEP-$t$, the dominant source of the parametric uncertainty is $\Delta\alpha_{\rm had}^{(5)}(M_Z^2)$. In Sec.~\ref{sec:todo}, we will examine how improvement of each observable's precision affects the sensitivity to new physics.

\section{To Do List for a Successful Electroweak Program}
\label{sec:todo}

\begin{figure}[!h]\begin{center}
\includegraphics[width=0.38\textwidth]{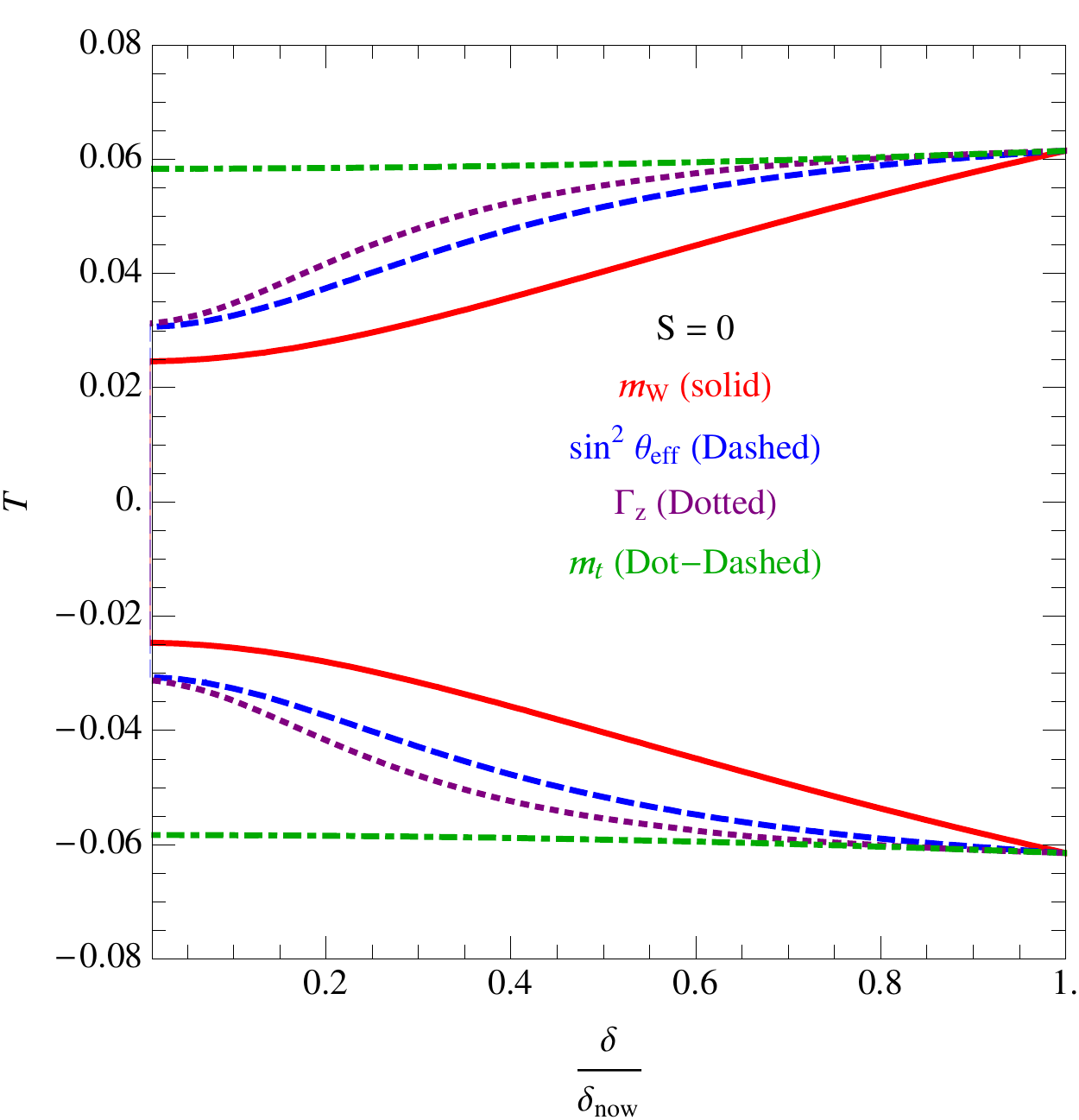}  \quad \includegraphics[width=0.38\textwidth]{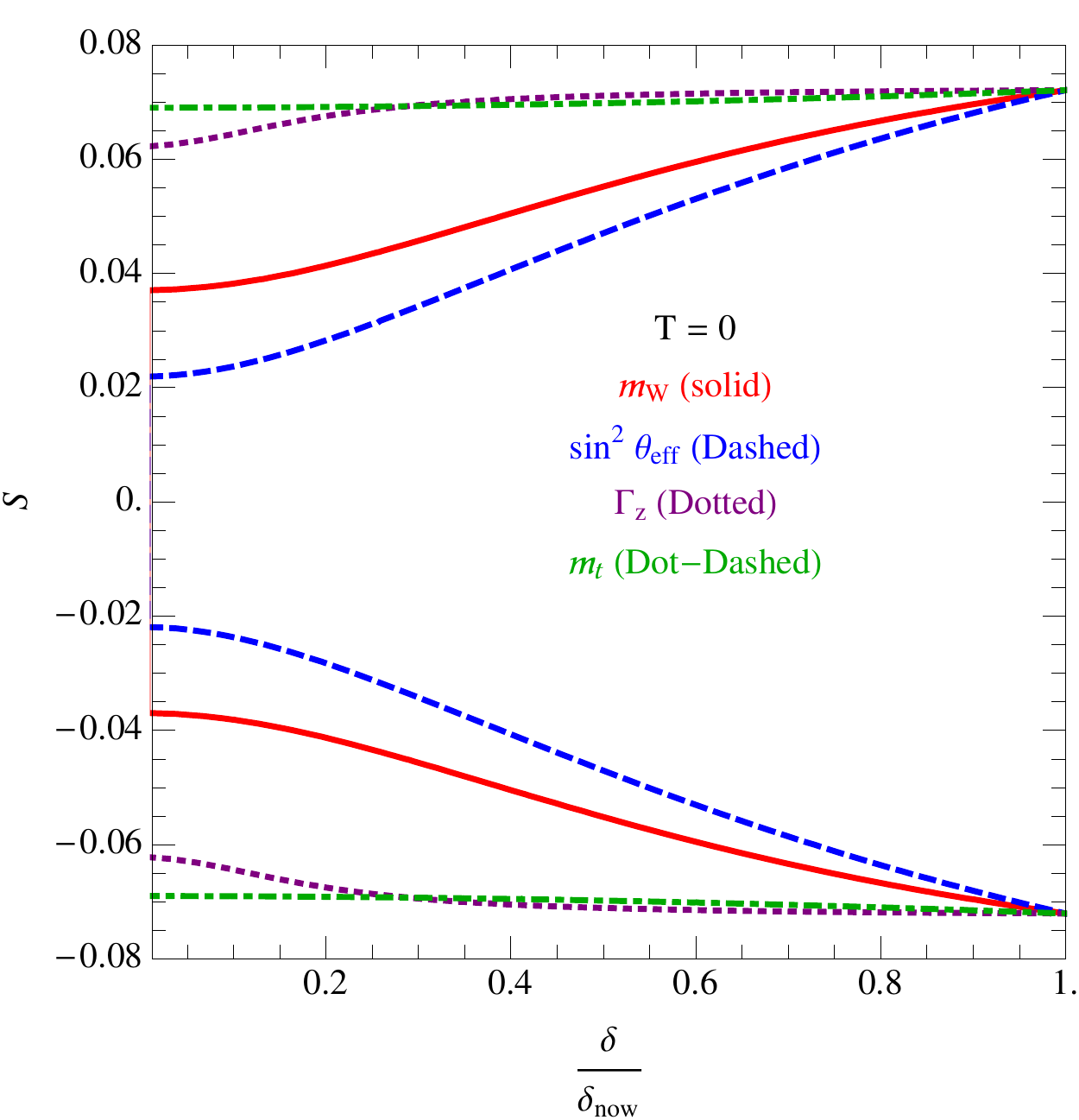}\\
\includegraphics[width=0.38\textwidth]{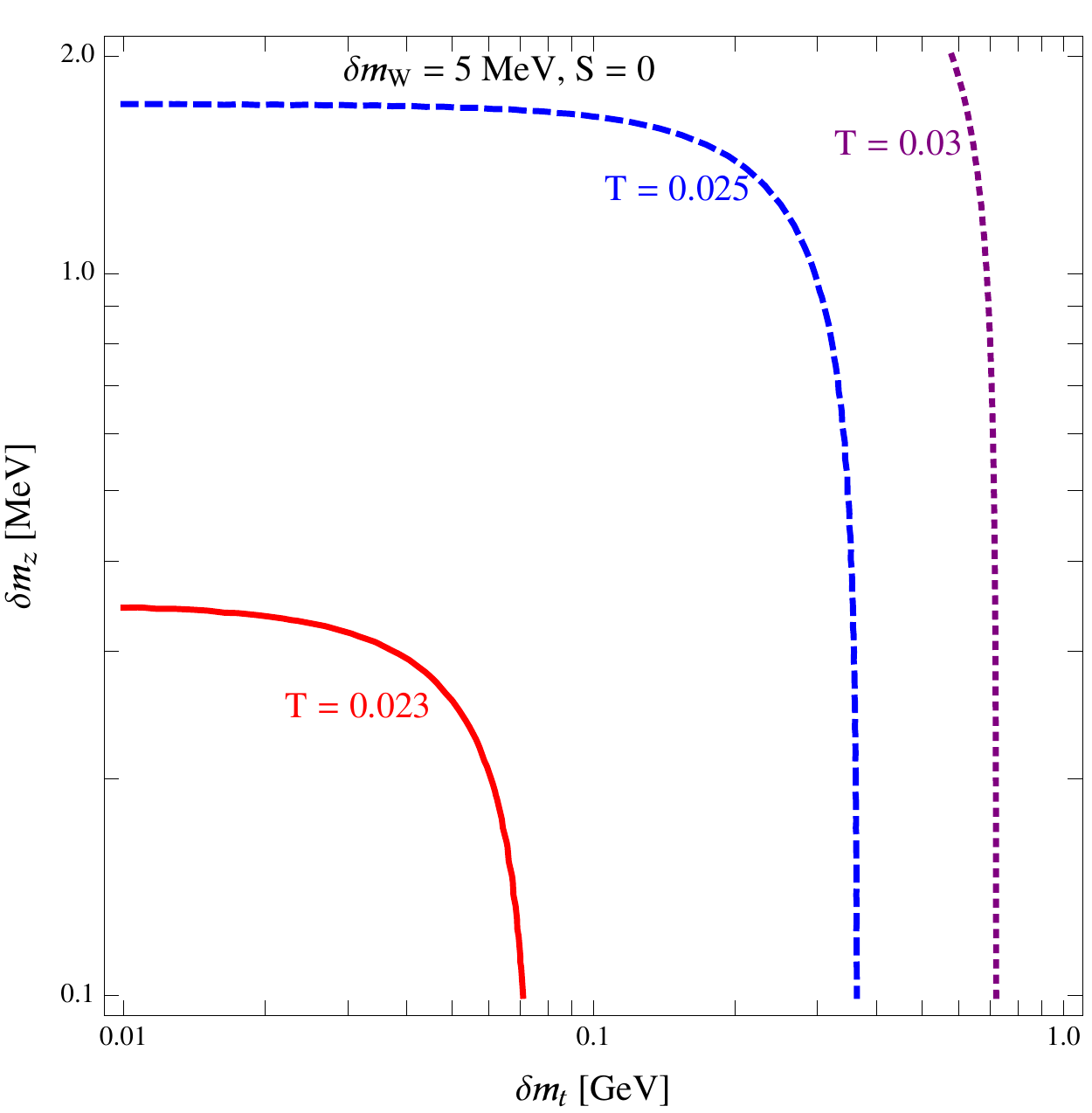}\quad  \includegraphics[width=0.38\textwidth]{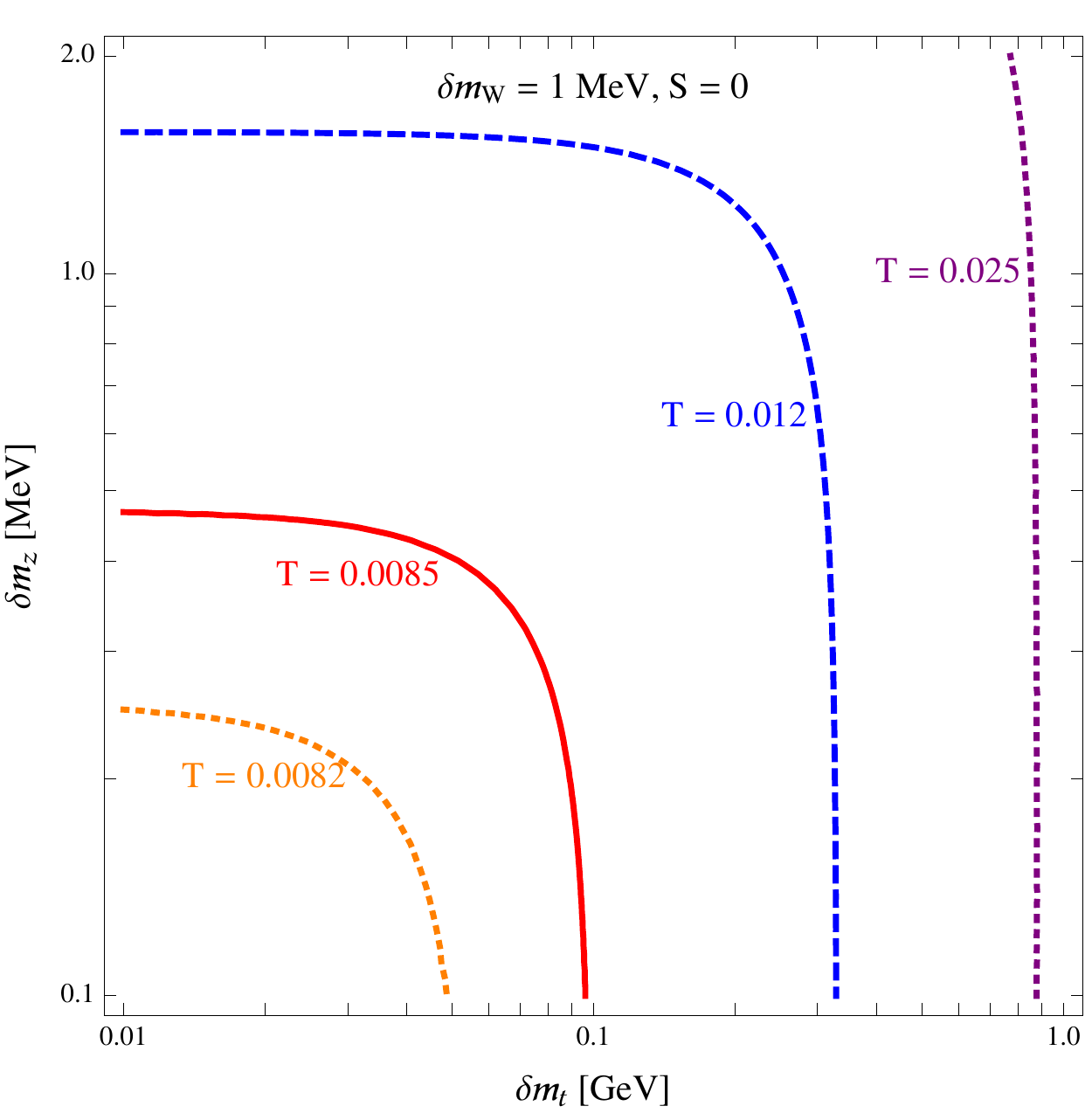}   \\
\includegraphics[width=0.38\textwidth]{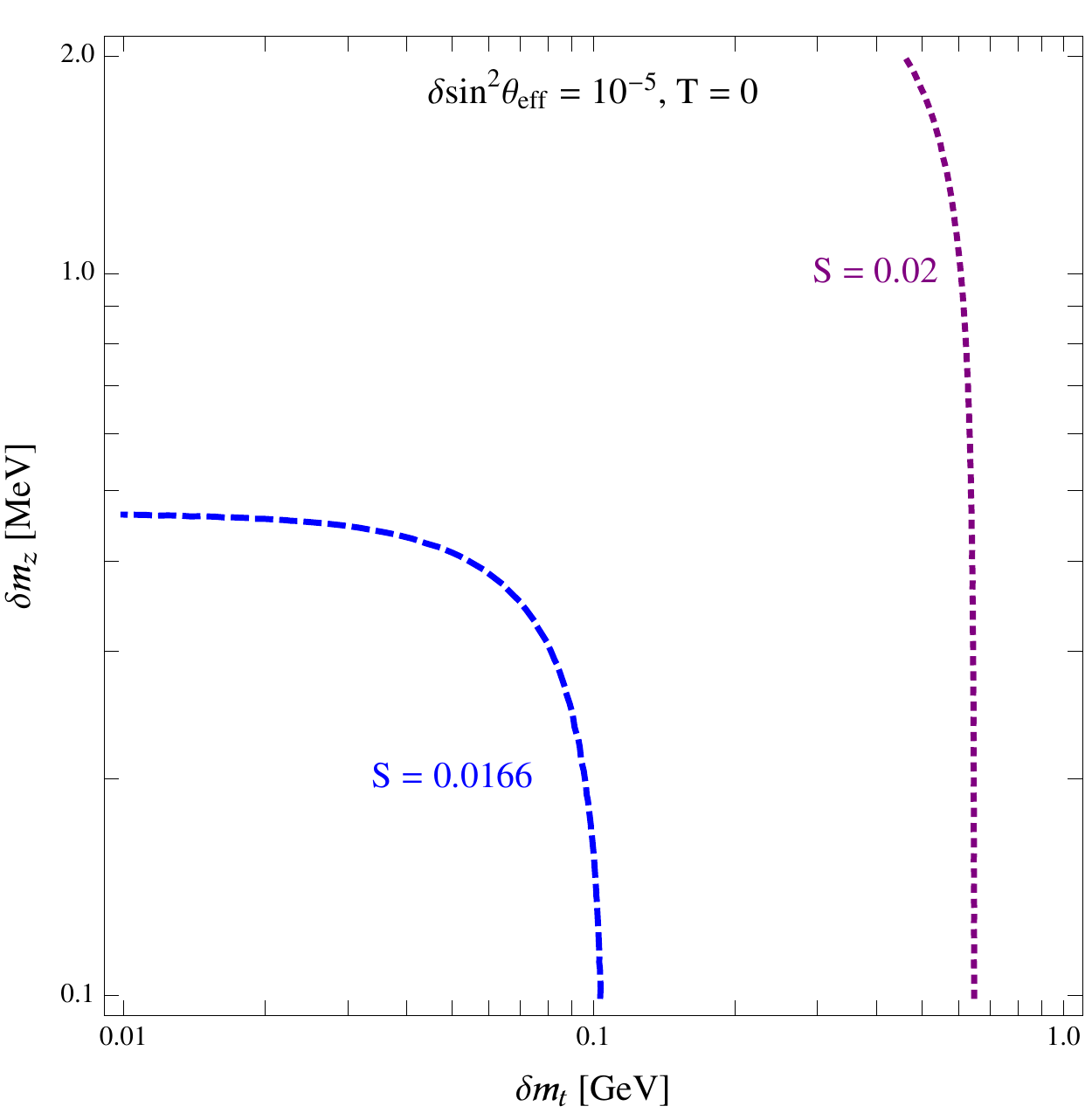}   \quad \includegraphics[width=0.41\textwidth]{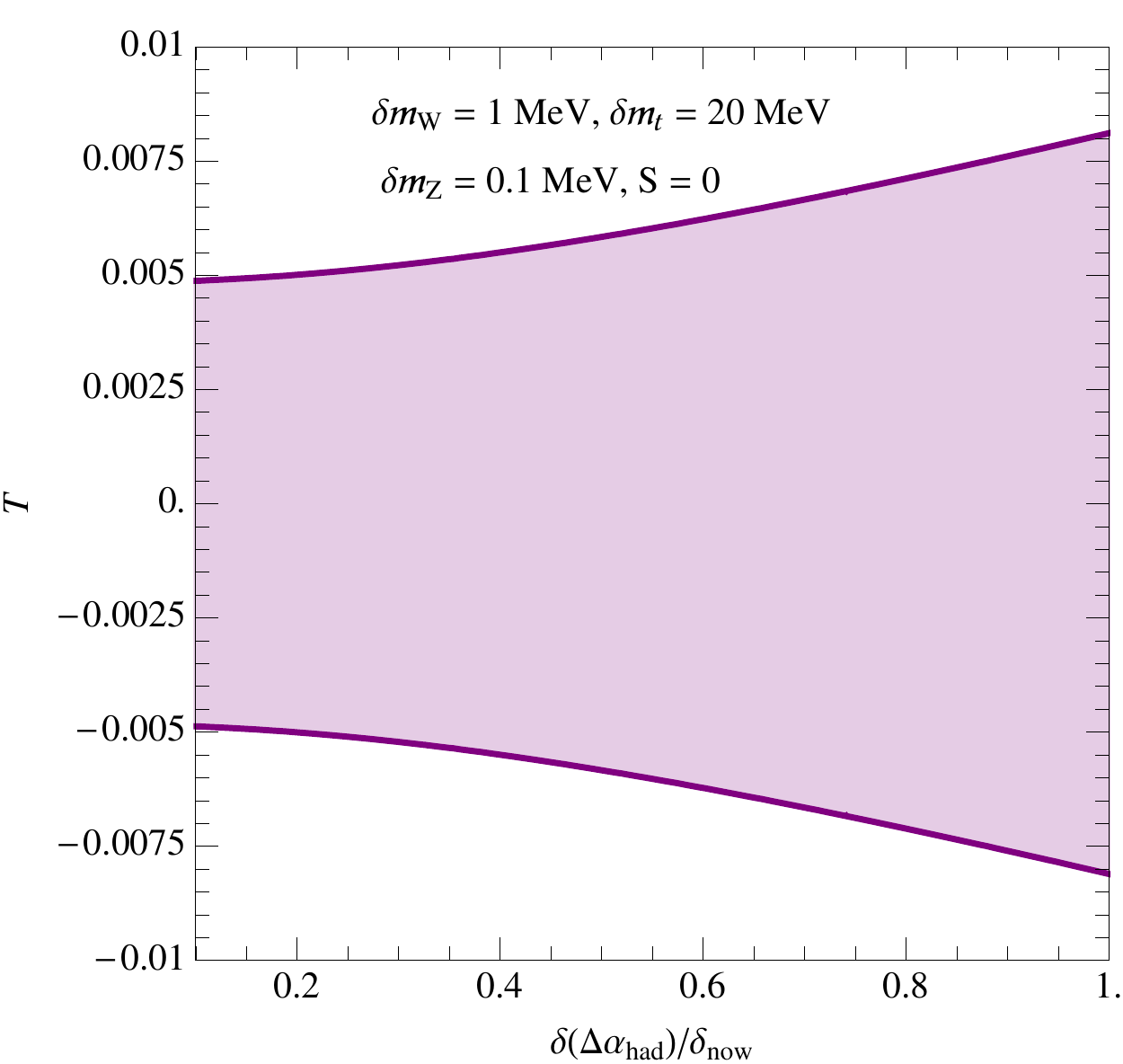}\end{center}
\caption{First row: allowed $T$ (left) and $S$ (right) at 2$\sigma$ C.L. as a function of error bar of one observable (normalized with respect to its current value) with the precisions of all the other observables in the fit fixed at current values. Second row: contours of allowed $T$ at 2 $\sigma$ C.L. in the $(\delta m_t, \delta m_Z)$ plane for $\delta m_W = $ 5 MeV (left) and 1 MeV (right). Again the precisions of all other observables in the fit fixed at current values. Last row: left plot: contours of allowed $S$ at 2$\sigma$ C.L. in the $(\delta m_t, \delta m_Z)$ plane for $\delta \sin^2\theta_{\rm eff}^\ell = 10^{-5}$ (left) ; right plot: allowed $T$ at 2$\sigma$ C.L. as a function of the error bar of $\Delta \alpha_{\rm had}^{(5)}$ normalized to its current value fixing $\delta m_W = 1$ MeV, $\delta m_t$ = 20 MeV and $\delta m_Z$ = 0.1 MeV.  }
\label{fig:sensitivity}
\end{figure}%

So far we have studied the reach of future $e^+e^-$ colliders for new physics parametrized by $S$ and $T$, based on estimated precisions of electroweak observables in the literature. In this section, we want to answer slightly different questions: what are the most important observables whose precisions need to be improved to achieve the best sensitivity of EWPT? What levels of precision are desirable for these observables? The answers are already contained in the simplified fits for different experiments but we want to make it clearer by decomposing the fit into three steps and changing the error bar of only one or two observables at each step. For this section, we will consider two limits with $S = 0$ or $T = 0$ and consider only the bound on $T$ or $S$. 

Among all electroweak observables, $m_W$ is the one that is most sensitive to the $T$ parameter and $\sin^2\theta_{\rm eff}^\ell$ is the one most sensitive to the $S$ parameter. This is demonstrated by the plots in the first row of Fig.~\ref{fig:sensitivity}, where we presented the dependence of $T$ setting $S=0$ (left panel) and $S$ setting $T=0$ (right panel) on four observables: $m_W$, $\sin^2\theta_{\rm eff}^\ell$, $\Gamma_Z$ and $m_t$.  Keeping the other observables with the current precisions, the allowed $T$ at 2$\sigma$ C.L. will decrease by a factor of 2 if the $m_W$ error bar is reduced from the current value 15 MeV to 5 MeV, the ILC projection. This is actually the main source of improvement for $T$ at ILC over LEP. The allowed $T$ at 2$\sigma$ C.L. could be reduced by a factor of 3 if the $m_W$ error bar is reduced to about a few hundred keV to 1 MeV, the TLEP-$W$ projection. This reduction could also be achieved if $\sin^2 \theta_{\rm eff}^\ell$ and/or $\Gamma_Z$ could be measured with errors of $2\times 10^{-5}$ and/or 200 keV respectively. This explains why the sensitivity of TLEP-$Z$ and TLEP-$W$ are almost exactly the same in terms of constraining $S$ and $T$. TLEP-$Z$ could measure the weak mixing angle and the $Z$ width very precisely and improving $m_W$ precision {\em only} does not help improve the sensitivity further.
Thus the priority of all electroweak programs is to improve the measurements of $m_W$ or $\sin^2\theta_{\rm eff}^\ell$ and reduce their theory uncertainties as well. 

For $m_W$ as well as the other derived observables, the errors of $m_t$ and $m_Z$ are the dominant sources of parametric uncertainties at the moment as is demonstrated in Table~\ref{tab:errors}. Thus among all free observables in the fit, $m_t$ and $m_Z$ are the most important ones to improve the sensitivity to new physics further. The effect on $T$ from reducing the error bars of $m_t$ and $m_Z$ for different choices of $\delta m_W$ is presented in the middle row of Fig.~\ref{fig:sensitivity}. In these two plots, we fix the errors of all the other observables in the fit to their current values. For $\delta m_W$ around or above 5 MeV, improving $\delta m_t$ and $\delta m_Z$ doesn't help much. When $\delta m_W$ drops to around 1 MeV, reducing $\delta m_Z$ by at least a factor of 4 and $\delta m_t$ by at least a factor of 10 compared to their current values {\em simultaneously} could improve the constraint on $T$ by a factor of about 3. This explains that TLEP-$t$ could improve the sensitivity to new physics by a factor of 10 compared to the current constraint along the $T$ axis with a factor of 3 from shrinking $\sin^2\theta_{\rm eff}^\ell$ and $\delta m_W$ and another factor of 3 from simultaneous reductions in $\delta m_t$ and $\delta m_Z$. However, along the $S$ axis, reducing $\delta m_t$ and $\delta m_Z$ doesn't help much as depicted in the right panel of the bottom row in Fig.~\ref{fig:sensitivity}.  

Lastly once $\delta m_t$ is reduced to be below 100 MeV and $m_Z$ is reduced to be below 0.5 MeV, they are no longer the dominant sources of parametric uncertainties while the contribution from $\Delta \alpha_{\rm had}^{(5)}$ will become the most important one. The improvement of $T$ as a function of the error bar of $\Delta \alpha_{\rm had}^{(5)}$ is depicted in the last row of Fig.~\ref{fig:sensitivity} fixing $\delta m_W = 1$ MeV, $\delta m_t$ = 20 MeV and $\delta m_Z$ = 0.1 MeV. Reducing the error bar of $\Delta \alpha_{\rm had}^{(5)}$ by a factor of 5 or more may only buy us a mild improvement of allowed $T$ range about 2. 

In summary, the following observables are the most important ones for EWPT and they should be determined with precisions 
\begin{itemize}
\item Determine $m_W$ to better than 5 MeV precision and $\sin^2 \theta_{\rm eff}^\ell$ to better than $2 \times 10^{-5}$ precision . 
\item Determine $m_t$ to 100 MeV precision and $m_Z$ to 500 keV precision.
\end{itemize}
Notice that in the discussions of this section, we do not differentiate theory uncertainties from experimental ones. It should be understood that the precision goals apply to both experimental and theory uncertainties. This means that for $m_W$ and $\sin^2\theta_{\rm eff}^\ell$, complete three-loop SM electroweak corrections computations are desirable. 

\section{Higgs Measurements at CEPC}
\label{sec:CEPChiggs}
\afterpage{\clearpage}

We have discussed the reach of CEPC measurements near the $Z$ pole for electroweak precision observables, but the main goal of CEPC is to perform high-luminosity measurements of Higgs boson properties. In this section we will provide a simple estimate of the expected precision of Higgs coupling measurements at CEPC. We do this by rescaling ILC estimates from the ILC Higgs White Paper~\cite{Asner:2013psa}. Table 5.4 of that paper presents a set of results for (among other scenarios) a 250 GeV ILC run accumulating 250 fb$^{-1}$ of data with polarized beams. At CEPC, the plan currently being discussed is to accumulate 5 ab$^{-1}$ of $\sqrt{s} = 240$ GeV data over 10 years, without polarized beams. At both the ILC and CEPC, the measurements of Higgs properties are expected to be dominated by statistical, rather than systematic, uncertainties. As a result, we can simply rescale the ILC's 250 GeV, 250 fb$^{-1}$ numbers to obtain CEPC uncertainties: $\Delta_{\rm CEPC} \approx \sqrt{\frac{\sigma_{\rm ILC}/\sigma_{\rm CEPC}}{{\cal L}_{\rm CEPC}/{\cal L}_{\rm ILC}}} \Delta_{\rm ILC}$. The luminosity ratio is ${\cal L}_{\rm CEPC}/{\cal L}_{\rm ILC} = 20$. The cross sections will differ for two reasons: first, CEPC plans to run at a center-of-mass energy of 240 GeV rather than 250 GeV. Second, CEPC is planning to run with unpolarized beams, while the ILC numbers quoted in ref.~\cite{Asner:2013psa} assume $P_{e^-} = -0.8$ and $P_{e^+} = +0.3$. We have computed the ratio of leading-order cross sections with appropriate beam polarizations using MadGraph~\cite{Stelzer:1994ta,Maltoni:2002qb,Alwall:2007st,Alwall:2011uj}. We find that:
\beq
\frac{\sigma_{\rm ILC}^{WW}(\nu {\overline \nu} h)}{\sigma_{\rm CEPC}^{WW}(\nu {\overline \nu} h)} & \approx & 2.89, \nonumber \\
\frac{\sigma_{\rm ILC}(Z h)}{\sigma_{\rm CEPC}(Z h)} & \approx & 1.44.
\eeq
The superscript $WW$ emphasizes that we are considering only the $\nu {\overline \nu} h$ contribution that does not go through an on-shell $Z$ boson (interference effects are small because the $Z$ is narrow). Thus, both cross sections are smaller at CEPC, and the case of $e^- e^+ \to \nu {\overline \nu} h$ production through $WW$ fusion is significantly smaller. As a result, the uncertainties for the $Zh$ process scale as $\Delta_{\rm CEPC} \approx 0.26 \Delta_{\rm ILC}$ whereas for the case of $WW$ fusion, $\Delta_{\rm CEPC} \approx 0.38 \Delta_{\rm ILC}$. These resulting uncertainties are displayed in the left-hand part of Table~\ref{tab:CEPChiggs}.

\begin{table}[h]
\begin{center}
\setlength{\tabcolsep}{.3em}
\begin{tabular}{|c|c|c|}
\hline
$\sqrt{s}$ and ${\cal L}$ & \multicolumn{2}{c|}{CEPC: 5 ab$^{-1}$, 240 GeV} \\
\hline
& $Zh$ & $\nu {\overline \nu} h$ \\
\hline
$\Delta \sigma/\sigma$ & 0.70\% & - \\
\hline
mode & \multicolumn{2}{c|}{$\Delta(\sigma \cdot {\rm Br})/(\sigma \cdot{\rm Br})$} \\
\hline
$h \to b{\overline b}$ & 0.32\% & 4.0\% \\
$h \to c{\overline c}$ & 2.2 \% & - \\
$h \to gg$ & 1.9\% & - \\
$h \to WW^*$ & 1.7\% & - \\
$h \to \tau^+ \tau^-$ & 1.1\% & - \\
$h \to ZZ^*$ & 4.8\% & - \\
$h \to \gamma\gamma$ & 9.1\% & - \\
$h \to \mu^+\mu^-$ & 27\% & - \\
\hline
\end{tabular}
\quad
\begin{tabular}{|c|c|c|}
\hline
Coupling & CEPC (5 ab$^{-1}$) & CEPC + HL-LHC \\
\hline
$\gamma\gamma$ & 4.8\% & 1.7\% \\
$gg$ & 1.9\% & 1.8\% \\
$WW$ & 1.6\% & 1.6\% \\
$ZZ$ & 0.20\% & 0.20\% \\
$t {\overline t}$ & 1.9\% & 1.9\% \\
$b {\overline b}$ & 1.5\% & 1.5\% \\
$\tau^+ \tau^-$ & 1.7\% & 1.6\% \\
\hline
\end{tabular}
\caption{Estimated uncertainties in Higgs measurements at CEPC. At left: uncertainties in cross section and cross section times branching ratio measurements, analogous to Table 5.4 in the ILC Higgs White Paper~\cite{Asner:2013psa}. At right: uncertainties on individual Higgs couplings from a profile likelihood in a seven parameter fit, analogous to Table 6.4 of ref.~\cite{Asner:2013psa}. The third column includes a 3.6\% constraint on the ratio ${\rm Br}(h \to \gamma\gamma)/{\rm Br}(h \to ZZ^*)$ from the high-luminosity LHC run~\cite{Peskin:2013xra}.}  
\label{tab:CEPChiggs}
\end{center}
\end{table}

Given the set of ten measurements in Table~\ref{tab:CEPChiggs}, we would like to know how well CEPC would constrain individual couplings of the Higgs boson to different particles. To answer this question we peform a seven-parameter $\chi^2$ fit for rescaling couplings by factors $\kappa_\gamma$, $\kappa_g$, $\kappa_W$, $\kappa_Z$, $\kappa_t$, $\kappa_b$, and $\kappa_\tau$. In this fit we assume that the up-type scaling factors are equal ($\kappa_c = \kappa_t$), the down-type scaling factors are equal, and the leptonic scaling factors are equal ($\kappa_\mu = \kappa_\tau$). This fit omits the interesting possibility of invisible or exotic decays~\cite{Curtin:2013fra}. Once we have constructed the $\chi^2$ as a function of the seven parameters, there are various choices we could make about what we mean by the $1\sigma$ error bar on each individual parameter. We choose a profile likelihood. To set a $1\sigma$ bound on $\kappa_\gamma$, for instance, we find the value of the other six $\kappa$ parameters that minimizes the $\chi^2$:
\beq
\chi^2(\kappa_\gamma) \equiv \min_{\kappa_g,\ldots \kappa_\tau} \chi^2(\kappa_\gamma, \kappa_g, \ldots \kappa_\tau).
\eeq
We then look for the value at which $\Delta \chi^2(\kappa_\gamma) = 1$ to set the 68\% CL limit. We have checked that performing this procedure on the ILC measurement uncertainties in Table 5.4 of ref.~\cite{Asner:2013psa} reproduces the $\kappa$ constraints in Table 6.4 of the same reference. An alternative procedure would be to {\em marginalize} over the other six $\kappa$ parameters by integrating the likelihood with a flat prior, as in ref.~\cite{Peskin:2012we}. Such a procedure yields similar results, with slightly less conservative bounds. (Ref.~\cite{Peskin:2012we} also imposed the constraint that $\kappa_W$ and $\kappa_Z$ are $\leq 1$, a theoretically well-motivated procedure which we choose not to do for consistency with the results of ref.~\cite{Asner:2013psa}.) In making these estimates, we have ignored theory uncertainties, which were taken to be 0.1\% in ref.~\cite{Asner:2013psa}. This is sufficiently small as to make little difference in the fit. A detailed discussion of how lattice QCD can reduce the relevant theory uncertainties may be found in ref.~\cite{Lepage:2014fla}, which concludes that theory uncertainties can be made small enough that experimental uncertainties dominate for Higgs coupling determination. In the final column of Table~\ref{tab:CEPChiggs} at right, we also show the combination with the LHC's constraint on the ratio of Higgs decay widths to photons and $Z$ bosons. This is expected to be measured to a precision of 3.6\% with small theoretical uncertainty~\cite{Peskin:2013xra}. Combining with this information significantly improves CEPC's constraint on the Higgs coupling to photons, but has little effect on the precision with which other couplings can be extracted.

\section{New Physics Reach and Complementarity}
\label{sec:newphysics}

Precision $Z$ and $W$ boson measurements and precision Higgs boson measurements both offer the possibility to probe new physics at energy scales out of direct reach. They are sensitive to different operators. For instance, the $T$ parameter operator $\left|h^\dagger D_\mu h\right|^2$ is highly constrained by measurements of the $W$ mass and $\sin^2 \theta^\ell_{\rm eff}$, while Higgs coupling measurements are sensitive to operators like $\partial_\mu (h^\dagger h) \partial^\mu (h^\dagger h)$ and $h^\dagger h B_{\mu \nu} B^{\mu \nu}$. Different models of new physics make different predictions for the size of these operators, and so in the event that new physics is within reach it could be important to have the full suite of precision electroweak and Higgs measurements as a ``fingerprint'' for the new physics.

On the other hand, in many models the predictions for different observables are correlated, so we can make model-independent comparisons of the reach for $S$ and $T$ parameter fits versus Higgs coupling measurements. In a companion paper, we will take a detailed look at how these measurements constrain natural SUSY theories with light stops and Higgsinos~\cite{Fan:2014axa}. For now, we will look at two simplified classes of new physics models. The first are composite Higgs theories in which the Higgs is a pseudo-Nambu-Goldstone boson arising from the breaking of a global symmetry extending the electroweak group, the relevant properties of which are reviewed in refs.~\cite{Contino:2010rs,Azatov:2012qz}. The second is the case of SUSY as represented by a left-handed stop, with other particles decoupled.

If the Higgs boson is composite, there will be a plethora of new states that play a role in electroweak symmetry breaking, and the Higgs alone will not fully unitarize $W$ and $Z$ boson scattering. This means that the Higgs coupling to $W^+W^-$ and $ZZ$ final states is modified on the order of $v^2/f^2$, where $f$ is the decay constant for the PNGB Higgs. For example, in the minimal composite Higgs model \cite{Agashe:2004rs}, we have:
\beq
\kappa_W = \kappa_Z = \sqrt{1 - \frac{v^2}{f^2}},
  \label{eq:hVVcompositecorrection}
\eeq
 Because the primary Higgs production mechanism at an $e^+ e^-$ collider is Higgsstrahlung, $e^+ e^- \to Z^* \to Z h$, the coupling $\kappa_Z$ is especially well-measured and provides a powerful constraint on the scale $f$. The details of how a composite Higgs theory modifies the $S$ and $T$ parameters are model-dependent. As a general guideline they receive corrections suppressed by the scale $m_\rho$, the mass of a technirho meson, i.e. a composite state sourced by the SU(2)$_L$ current. We expect contributions to the $S$ parameter of order
\beq
S \sim \frac{4 \pi v^2}{m_\rho^2} \sim \frac{N}{4\pi} \frac{v^2}{f^2},
\eeq 
where we have used the NDA estimate $m_\rho \sim 4\pi f/\sqrt{N}$. The number of colors $N$ in the composite sector is generally order one---rarely larger than 10 due to phenomenological constraints like Landau poles and cosmological problems---and so we will take as our benchmark estimate
\beq
S \approx \frac{v^2}{4 f^2}. 
   \label{eq:Scompositeestimate}
\eeq
Comparing equations~\ref{eq:hVVcompositecorrection} and~\ref{eq:Scompositeestimate}, we see that the parametric size of corrections to Higgs boson couplings and to the $S$ parameter are linked.

In the case of SUSY, we consider left-handed stops. Their dominant effect on Higgs couplings is to run in the loop coupling the Higgs to gluons:
\beq
\kappa_g -1 \approx \frac{m_t^2}{4 m_{{\tilde t}_L^2}}.
\eeq
They also modify the photon coupling $\kappa_\gamma$ by a smaller amount, which we will ignore for the moment (but include in the companion paper). The dominant effect of stops on the $S$ and $T$ parameters is to induce a contribution to $T$~\cite{Drees:1990dx}:
\beq
T \approx \frac{m_t^4}{16 \pi \sin^2 \theta_W m_W^2 m_{{\tilde t}_L}^2}.
\eeq
There is a small negative contribution to the $S$ parameter that we ignore for now.

\begin{table}[!h]
\centering
\setlength{\tabcolsep}{.3em}
\begin{tabular}{|c|c|c|c|c|}
\hline
Experiment & $\kappa_Z$ (68\%) &  $\  f$ (GeV)& $\kappa_g$ (68\%) & $m_{{\tilde t}_L}$ (GeV) \\
\hline
HL-LHC & 3\% & 1.0 TeV & 4\% & 430 GeV \\
ILC500 & 0.3\% & 3.1 TeV & 1.6\% & 690 GeV \\
ILC500-up & 0.2\% & 3.9 TeV & 0.9\% & 910 GeV \\
CEPC & 0.2\% & 3.9 TeV & 0.9\% & 910 GeV \\
TLEP & 0.1\% & 5.5 TeV & 0.6\% & 1.1 GeV \\
\hline
\end{tabular}
\caption{Interpreting Higgs coupling bounds in terms of new physics reach.}  
\label{tab:kappainterp}
\end{table}

\begin{table}[!h]
\centering
\setlength{\tabcolsep}{.3em}
\begin{tabular}{|c|c|c|c|c|}
\hline
Experiment & $S$ (68\%) & $f$ (GeV) & $T$ (68\%) & $m_{{\tilde t}_L}$ (GeV) \\
\hline
ILC & 0.012 & 1.1 TeV & 0.015 & 890 GeV \\
CEPC (opt.) & 0.02 & 880 GeV & 0.016 & 870 GeV \\
CEPC (imp.) & 0.014 & 1.0 TeV & 0.011 & 1.1 GeV \\
TLEP-$Z$ & 0.013 & 1.1 TeV & 0.012 & 1.0 TeV \\
TLEP-$t$ & 0.009 & 1.3 TeV & 0.006 & 1.5 TeV \\
\hline
\end{tabular}
\caption{Interpreting $S$ and $T$ parameter bounds in terms of new physics reach. CEPC (imp.) is assuming the improvement in both $\sin^2 \theta_{\rm eff}^\ell$ and $\Gamma_Z$, as discussed in Section~\ref{subsec:cepc_imp}. }  
\label{tab:STinterp}
\end{table}

In Table~\ref{tab:kappainterp}, we present the relevant 1$\sigma$ error bars for the Higgs couplings $\kappa_Z$ and $\kappa_g$ for various experiments: we performed a one parameter fit with either $\kappa_Z (= \kappa_W)$ or $\kappa_g$. We also translate these into bounds on the scale $f$ in composite Higgs models and on the left-handed stop mass in SUSY models, respectively, to give some indication of how measurement accuracy translates to a reach for heavy particles. In Table~\ref{tab:STinterp}, we present the value of $S$ where the line $T = 0$ intersects the 68\% CL ellipse, and vice versa, from our calculation in Figs.~\ref{fig:ST} and~\ref{fig:cepc}. We also translate these into bounds on $f$ and on $m_{{\tilde t}_L}$, respectively. Of course, bounds on new physics are always model-dependent and the relative sizes of various operators will depend on the model. Here we can see that for a composite Higgs, the most powerful probe is the very well-measured coupling of the Higgs to the $Z$ boson. The bounds from this measurement dwarf those from the $S$ and $T$ parameters. On the other hand, bounds on the left-handed stops from the $T$ parameter and from Higgs coupling measurements are very similar, with the $T$ parameter bound generally being slightly stronger. This points to an important complementarity between Higgs factory measurements and $Z$ factory (or $W$ and top threshold) measurements. Both sets of measurements are crucial to obtain a broad view of what possible new electroweak physics can exist at the TeV scale.

We have treated the Higgs measurements independently of the $(S,T)$ plane fits to illustrate the new physics reach of different observables. However, they are related: for example, the $S$ parameter operator $h^\dagger \sigma^i h W^i_{\mu \nu} B_{\mu \nu}$ modifies the partial widths for Higgs boson decays to two electroweak bosons. The proper procedure once all the data is available will be to do a global fit combining all known pieces of information.

\section{Conclusions}
\label{sec:conclusion}
In this paper we perform a global fit of electroweak observables with oblique corrections and estimate the size of the region in the $(S, T)$ plane that will be allowed by several future high-precision measurements: the ILC GigaZ program, the FCC-ee TeraZ program, extended runs of FCC-ee combining $Z$ pole data with data at the $W^+ W^-$ threshold and the $t{\overline t}$ threshold, and the $Z$ pole program of CEPC. In particular, the reach of CEPC for new physics that could be parametrized by oblique parameters is presented for the first time. We also discuss possible ways to improve the CEPC baseline program. Compared to current sensitivity, the ILC and CEPC baseline programs could improve the sensitivity to new physics encoded in $S$ and $T$ by a factor of $\sim 3$ while the FCC-ee program and proposed improved CEPC measurements could improve by a factor $\sim 10$. 
We also discuss many of the relative advantages and disadvantages of the different machines; for example, the $Z$ mass measurement will be improved only at circular colliders, which can follow LEP in exploiting resonant spin depolarization. We emphasize the basic physics of the fits and their potential bottlenecks, specifying the goals of the electroweak program in future colliders in order to achieve the best sensitivity. For example, given current data the highest priorities are reducing the uncertainties on $m_W$ for determination of $T$ and of $\sin^2 \theta_{\rm eff}$ for determination of $S$, while improved measurements of the top quark mass or the hadronic contribution to the running of $\alpha$ become important only once other error bars have been significantly reduced. In addition, we perform a first seven-parameter fit of Higgs couplings to demonstrate the power of the CEPC Higgs program and study the complementarity between future electroweak precision and Higgs measurements in probing new physics scenarios such as natural supersymmetry and composite Higgs.

\section*{Acknowledgments}
We thank Weiren Chou, Ayres Freitas, Paul Langacker, Zhijun Liang, Xinchou Lou, Marat Freytsis, Matt Schwartz, Witek Skiba and Haijun Yang for useful discussions and comments. We thank the CFHEP in Beijing for its hospitality while this work was initiated and a portion of the paper was completed. The work of MR is supported in part by the NSF Grant PHY-1415548. L-TW is supported by the DOE Early Career Award under Grant DE-SC0003930. 

\appendix

\section{Treatment of Theory Uncertainties}
\label{app:theory}

\begin{figure}[!h]\begin{center}
\includegraphics[width=0.6\textwidth]{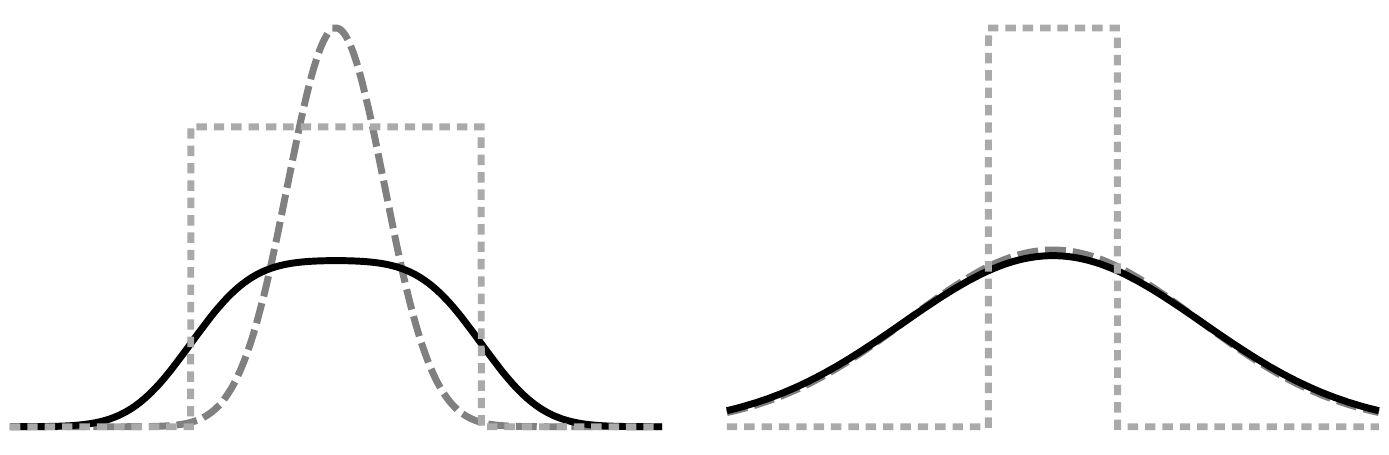} 
\end{center}
\caption{Illustration of the likelihood functions (solid black curves) that arise from convolving a Gaussian experimental uncertainty (dashed gray curves) and a flat theory uncertainty (dotted gravy curves), as in eq.~\ref{eq:convolution}. If the theory uncertainty is relatively small, as in the right-hand case ($\delta = 0.4, \sigma = 0.9$), it has little effect. If it is large, as in the left-hand case ($\delta = 0.9, \sigma = 0.3$), it stretches the peak of the Gaussian out into a flat plateau.}
\label{fig:convolution}
\end{figure}%

Uncertainties in fitting the theory to data arise not only from experimental measurement systematics and statistical fluctuations, but from theoretical uncertainties in relating the underlying parameters to observables. We include theory uncertainties in a similar manner to refs.~\cite{Hocker:2001xe,Flacher:2008zq,Lafaye:2009vr}. For instance, the measured top mass $m^{\rm meas}_t = 173.34 \pm 0.76$ GeV gives an experimental error bar on a parameter we can loosely refer to as the top quark mass~\cite{ATLAS:2014wva}. However, the fundamental top mass parameter $m^{\rm theory}_t$ (defined, for instance, in the $\overline{\rm MS}$ scheme or the 1S scheme), which we might use an input in computing other observables, is related to $m^{\rm meas}_t$ only up to some uncertainty of order a GeV. There is no particular reason to think that this uncertainty is Gaussian. Instead, we take theory uncertainties to be flat over some range and zero elsewhere. Given a fundamental set of theory parameters $\alpha_i$, we imagine that each observable $O_j$ is determined by theory to take a value $O^{\rm pred}_j(\alpha_1, \ldots \alpha_n)$ only up to some uncertainty $\delta_j$:
\beq
p(O_j | \alpha_1, \ldots \alpha_n) = \left\{
	\begin{array}{ll}
		\frac{1}{2 \delta_j} & \mbox{if } \left|O_j - O_j^{\rm pred}(\alpha_1, \ldots \alpha_n)\right| \leq \delta_j \\
		0 & \mbox{if } \left|O_j - O_j^{\rm pred}(\alpha_1, \ldots \alpha_n)\right| > \delta_j
	\end{array}
\right.
\eeq
Here by $O_j$ we mean the {\em true} value of the observable, assuming perfect measurement. On the other hand, the true value of an observable determines the measured value $M_j$ only up to some experimental precision $\sigma_j$, which we generally take to be Gaussian:
\beq
p(M_j | O_j) = \frac{1}{\sqrt{2\pi}\sigma_j} \exp\left(-\frac{\left(M_j - O_j\right)^2}{2\sigma_j^2}\right).
\eeq
From this, we extract the probability distribution (i.e., the likelihood) for given measurements in terms of fundamental theory parameters as a convolution, integrating out the unknown true value $O_j$ of the observable:
\beq
p(M_j | \alpha_1 \ldots \alpha_n) = \int dO_j p(M_j | O_j)p(O_j | \alpha_1, \ldots \alpha_n) = q\left(M_j ; O_j^{\rm pred}(\alpha_1, \ldots \alpha_n), \sigma_j, \delta_j\right),
\eeq
where
\beq
q(x; \mu, \sigma, \delta) \equiv \frac{1}{4\delta} \left({\rm erf}\left(\frac{x-\mu+\delta}{\sqrt{2}\sigma}\right) - {\rm erf}\left(\frac{x-\mu-\delta}{\sqrt{2}\sigma}\right)\right). \label{eq:convolution}
\eeq
This is, roughly speaking, a Gaussian that has been ``stretched'' so that its peak has width $\delta$, as illustrated in Fig.~\ref{fig:convolution}. If we normally defined a $\chi^2$ as $\left(x - \mu\right)^2/\sigma^2$, we can define a modified $\chi^2$ taking theoretical uncertainty into account as
\beq
\chi_{\rm mod}^2(x; \mu, \sigma, \delta) = -2 \log q(x; \mu, \sigma, \delta) - 2 \log(\sqrt{2\pi}\sigma).
\eeq
The second term plays no role in determining exclusion contours because they depend only on {\em differences} of $\chi^2$ values, but just ensures that this definition approaches the usual definition of $\chi^2$ as $\delta \to 0$.

\bibliography{ref}
\bibliographystyle{utphys}
\end{document}